\documentclass[aps,prl,preprint,superscriptaddress]{revtex4-2}

\usepackage{amsmath}
\usepackage{graphicx}
\usepackage{float}
\usepackage[version=4]{mhchem}
\usepackage{siunitx}
\DeclareSIUnit{\counts}{counts}
\DeclareSIUnit{\rpm}{rpm}
\usepackage[left]{lineno}

\usepackage{hyperref}
\usepackage[nameinlink]{cleveref}
\crefname{figure}{Fig.}{Figs.}
\Crefname{figure}{Fig.}{Figs.}
\crefname{equation}{Eq.}{Eqs.}
\Crefname{equation}{Eq.}{Eqs.}
\crefname{section}{Sec.}{Secs.}
\Crefname{section}{Sec.}{Secs.}

\begin{document}

\title{Single Best Fits Can Be Misleading: Resolving Common Trapping Signatures across FA–Cs Perovskites}

\author{Maxim Simmonds}
\email[]{maximsimmonds@gmail.com}
\affiliation{Helmholtz-Zentrum Berlin (HZB), Hahn-Meitner-Platz 1, 14109 Berlin, Germany}

\author{Katarzyna Pydzińska-Białek}
\email[]{katarzyna.pydzinska-bialek@helmholtz-berlin.de}
\affiliation{Helmholtz-Zentrum Berlin (HZB), Hahn-Meitner-Platz 1, 14109 Berlin, Germany}

\author{Thomas C. Rossi}
\email[]{thomas.rossi@helmholtz-berlin.de}
\affiliation{Helmholtz-Zentrum Berlin (HZB), Hahn-Meitner-Platz 1, 14109 Berlin, Germany}

\author{Mostafa Othman}
\email[]{mostafa.othman@epfl.ch}
\affiliation{École Polytechnique Fédérale de Lausanne (EPFL),
Rue de la Maladière 71b, Neuchâtel 2002, S witzerland}

\author{Aïcha Hessler-Wyser}
\email[]{aicha.hessler@epfl.ch}
\affiliation{École Polytechnique Fédérale de Lausanne (EPFL),
Rue de la Maladière 71b, Neuchâtel 2002, Switzerland}

\author{Christian Wolff}
\email[]{christian.wolff@epfl.ch}
\affiliation{École Polytechnique Fédérale de Lausanne (EPFL),
Rue de la Maladière 71b, Neuchâtel 2002, Switzerland}

\author{Christophe Ballif}
\email[]{christophe.ballif@epfl.ch}
\affiliation{École Polytechnique Fédérale de Lausanne (EPFL),
Rue de la Maladière 71b, Neuchâtel 2002, Switzerland}

\author{Vincent M. Le Corre}
\email[]{lecorre@mci.sdu.dk}
\affiliation{South Denmark University (SDU),
Alsion 2, Sønderborg, Denmark}

\author{Eva Unger}
\email[]{eva.unger@helmholtz-berlin.de}
\affiliation{Helmholtz-Zentrum Berlin (HZB), Hahn-Meitner-Platz 1, 14109 Berlin, Germany}

\date{\today}

\begin{abstract}
Identifying defects responsible for non-radiative recombination in metal-halide perovskites remains challenging, with no common defect signatures established across compositions. Here, we combine fluence-dependent time-resolved photoluminescence, full Shockley--Read--Hall modelling and Bayesian posterior inference across an FA$_{1-x}$Cs$_x$PbI$_3$ compositional series. We show that trap occupation governs carrier dynamics and defect-parameter identifiability in semiconducting materials: weakly occupied states exhibit electron-capture-coefficient--trap-density degeneracies, whereas trap filling lifts them. In our perovskite systems, a non-degenerate signature \(T\beta\), characterized by a strongly asymmetric \(\beta_p/\beta_n\) capture-coefficient ratio, occurs across all compositions, suggesting substantial trap filling is a general feature of perovskite carrier dynamics. We further identify \(T\alpha\) as a shared, device-performance-limiting non-radiative recombination channel, while a shallow \(T\epsilon\) signature unique to FAPbI$_3$ produces the largest steady-state non-radiative recombination rate and is consistent with the previously observed high stacking-fault density of this composition. This framework therefore provides a basis for comparing trapping signatures across different semiconducting materials at room temperature.
\end{abstract}

\maketitle

\section{Introduction}

Metal-halide perovskite thin films are promising solution-processable semiconductors for photovoltaics and other optoelectronic applications. However, their defects that govern device-limiting non-radiative recombination remain poorly understood~\cite{stranksJourneyGlobalUnderstanding2026a}. Developing a quantitative physical understanding of defect-mediated non-radiative recombination is therefore important for improving both the efficiency and operational stability of perovskite optoelectronic devices.

Direct identification of electrically active defects remains challenging because their concentrations are typically small relative to the atomic density of the material. Defects are therefore commonly inferred indirectly through their influence on carrier capture, emission and recombination dynamics. A range of time-resolved techniques are used to probe these dynamics in perovskite thin films, including transient absorption spectroscopy (TAS)~\cite{pydzinskaDeterminationInterfacialChargeTransfer2016}, optical-pump terahertz-probe spectroscopy (OPTP)~\cite{limLongrangeChargeCarrier2022}, time-resolved microwave conductivity (trMC)~\cite{guoModulatingNonradiativeRecombination2026} and time-resolved photoluminescence (trPL)~\cite{simmondsQuantifyingEvolvingDefect2026}. Among these, trPL is widely used because of its experimental simplicity and accessibility. Although trPL directly probes radiative emission, its temporal evolution is strongly influenced by competing non-radiative recombination and trapping processes.

Although trPL measurements are standard for perovskite thin films, their characteristically strong non-monoexponential behavior has challenged conventional descriptions of non-radiative recombination, in which the Shockley--Read--Hall (SRH) contribution is often represented by a single characteristic lifetime. Various phenomenological approaches, including ABC rate-equation models and single SRH lifetime-based descriptions, have been used to describe experimental transients~\cite{stranksNonradiativeLossesMetal2017,herzChargeCarrierDynamicsOrganicInorganic2016}. However, such models often fail to reproduce the measured kinetics consistently across broad ranges of carrier density~\cite{kiligaridisAreShockleyReadHallABC2021}. In particular, simplified lifetime-based descriptions, such as mono- or stretched-exponential models, cannot generally capture wide-dynamic-range trPL measurements, which exhibit power-law decays in regimes dominated by Shockley--Read--Hall recombination~\cite{yuanShallowDefectsVariable2024,yuanUnderstandingPowerLawPhotoluminescence2024}.

More recently, these limitations have motivated the development of models based on the full SRH formalism, in which electron/hole capture and emission, together with defect occupation, are treated explicitly rather than reduced to a single effective lifetime~\cite{shockleyStatisticsRecombinationHoles1952,simmonsNonequilibriumSteadyStateStatistics1971,yuanShallowDefectsVariable2024}. Within this framework, each trapping signature \(i\) is described by four parameters,
\[
\mathcal{P}_{i}
=
\left(
N_{t,i},\,\beta_{n,i},\,\beta_{p,i},\,E_{t,i}
\right),
\]
corresponding to its trap density (\(N_{t,i}\)), electron (\(\beta_{n,i}\)) and hole (\(\beta_{p,i}\)) capture coefficients, and trap energy (\(E_{t,i}\)), respectively. Here, the subscript \(i\) labels a single trapping signature, which we define as an electronically active state within the bandgap characterized by a unique parameter set \(\mathcal{P}_i\). A given structural defect may therefore give rise to multiple trapping signatures with distinct electronic properties. In this definition, no distinction is made a priori between shallow and deep trapping signatures, as in the case of perovskite thin films capture coefficients are predicted not to be correlated with trap energy~\cite{zhangDefectToleranceHalide2022}.

Collectively, full-SRH models have reproduced experimental kinetics over broad ranges of excitation fluence~\cite{yuanShallowDefectsVariable2024,kober-czernyDeterminingParametersMetalHalide2025,simmondsQuantifyingEvolvingDefect2026} and temperature~\cite{aalbersLowTemperaturesReduce2026}, and have shown that shallow defect states are compatible with the power-law photoluminescence decays observed experimentally~\cite{yuanShallowDefectsVariable2024,aalbersLowTemperaturesReduce2026}. However, these conclusions have largely been drawn from individual best-fit parameter sets~\cite{yuanShallowDefectsVariable2024,simmondsQuantifyingEvolvingDefect2026,aalbersLowTemperaturesReduce2026}.

The increased physical completeness of these models comes at the cost of a large number of fitted parameters that may be strongly correlated and non-unique. Consequently, excellent agreement with experiment does not imply that the inferred defect energies, densities and capture coefficients are uniquely constrained, and apparent differences between individual best-fit solutions may instead reflect different positions within correlated regions of parameter space. Such degeneracies are well established in the extraction of defect parameters from silicon absorbers, where combined temperature- and injection-dependent lifetime spectroscopy is commonly used to disentangle correlated defect properties~\cite{reinLifetimeSpectroscopyMethod2005}. Reliable comparison between samples therefore requires the joint uncertainty and correlation structure of the inferred parameters to be quantified.

Bayesian inference using sampling algorithms such as Markov chain Monte Carlo (MCMC) provides a route to quantifying these uncertainties and correlations. Such approaches have recently been applied to photoluminescence measurements of metal-halide perovskites~\cite{faiRapidOptoelectronicCharacterization2023,kober-czernyDeterminingParametersMetalHalide2025}, while MCMC-based inference has also been used to extract bulk and interface recombination parameters in III--V absorbers~\cite{grossmannGeneralizedModelingPhotoluminescence2023,renEmbeddingPhysicsDomain2020}. However, the identifiability of parameters in multi-defect full-SRH models remains poorly established. In particular, it remains unclear which parameters are identifiable from fluence-dependent perovskite trPL, how trap filling modifies correlations between inferred defect parameters, and whether posterior distributions can reveal recurring trapping signatures across different samples and compositions.

Here, we apply joint posterior inference to fluence-dependent trPL measurements of an FA$_{1-x}$Cs$_x$PbI$_3$ compositional series to identify effective trapping signatures and quantify the parameter degeneracies associated with their full-SRH descriptions.

Together, these results establish that posterior inference is a more robust framework for interpreting and comparing trapping signatures in perovskite thin films, with implications for understanding and ultimately mitigating defect-mediated recombination losses in photovoltaic and other semiconductor-based devices.

\section{Results}
\subsection{Sample selection and structural context}

The FA$_{1-x}$Cs$_x$PbI$_3$ films investigated here were newly fabricated following the same preparation procedure presented in~\textcite{othmanAlleviatingNanostructuralPhase2024}. This series exhibits a pronounced composition-dependent evolution in microstructure. In Cs-free FAPbI$_3$ ($x=0$), low-dose bright-field transmission electron microscopy and selected-area electron diffraction revealed a high density of stacking faults or $\{111\}_{\mathrm{C}}$ planar defects (see \cref{SI:TEM}). Their prevalence decreased with increasing Cs content and reached a minimum near $x=0.15$. At higher Cs concentrations, particularly for $x=0.30$, Cs-rich nonperovskite $\delta$-CsPbI$_3$ secondary phases were observed.

The intermediate composition therefore represents a regime in which both stacking faults and Cs-rich secondary phases are minimized. Consistent with this interpretation, macroscopic X-ray diffraction measurements reported by~\cite{othmanAlleviatingNanostructuralPhase2024} indicated the highest structural quality for FA$_{0.85}$Cs$_{0.15}$PbI$_3$. Based on these previously established microstructural trends, we selected the $x=0$, 0.15, and 0.30 compositions, hereafter denoted Cs0, Cs15, and Cs30, respectively, to investigate how nanoscopic structural defects affect resulting electronic trapping signatures.

These trends are consistent with broader reports that controlled Cs incorporation can modify the nanoscale structure of FA-rich perovskites. Cs-containing additives have been shown to reduce nanotwin and stacking-fault densities~\cite{phamUnravelingInfluenceCsCl2021}, while partial substitution of FA by Cs can modify quantum-confinement effects associated with nanoscale structural features~\cite{elmestekawyControllingIntrinsicQuantum2022}. Such phase impurities and structural heterogeneities are relevant because they have been associated with local optoelectronic losses and the initiation of degradation~\cite{macphersonLocalNanoscalePhase2022}.

\subsection{Optical characterization}

\begin{figure}[h!]
  \centering
  \includegraphics[width=1\columnwidth]{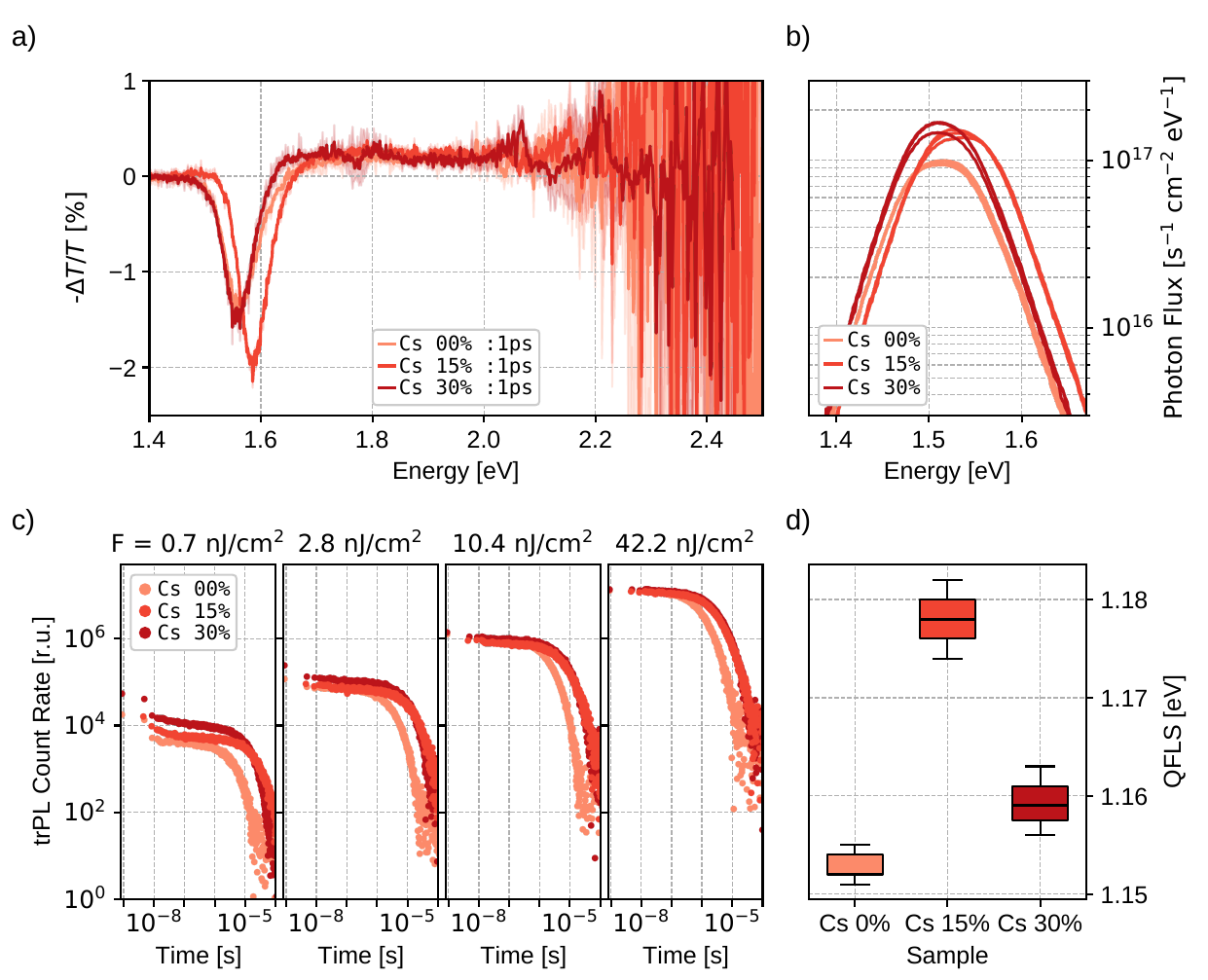}
  \caption{\label{fig:fig1}Steady-state and transient optical characterization of the FA$_{1-x}$Cs$_x$PbI$_3$ compositional series. (a) Transient-absorption spectra showing the ground-state-bleach positions of the three compositions at 1ps time delay and with a pump photon energy of \SI{1.77}{\electronvolt}. An additional weak higher-energy feature is observed for the Cs30 film near \SI{2.35}{\electronvolt}. Pump fluence of \SI{3.3}{\micro\joule\per\square\centi\metre} corresponding to ~\SI{3e17}{\per\cubic\centi\metre}. (b) Steady-state photoluminescence spectra for the Cs0, Cs15, and Cs30 films. The Cs15 film exhibits a shift of the PL maximum toward higher photon energies. (c) Fluence-dependent time-resolved photoluminescence measurements of the Cs0, Cs15, and Cs30 films over an excitation-fluence range of \SIrange[range-units = single]{0.7}{45}{\nano\joule\per\square\centi\metre}, corresponding to estimated photogenerated carrier densities of ~\SIrange[range-units=single]{6e13}{4e15}{\per\cubic\centi\metre} (d) Corresponding quasi-Fermi-level splittings extracted using the high-energy-tail fitting method. The Cs15 film exhibits the highest average QFLS. }
\end{figure}

First, we present the steady-state PL spectra in \cref{fig:fig1}b, and the corresponding quasi-Fermi-level splittings (QFLS) in \cref{fig:fig1}d, determined using the high-energy-tail fitting method~\cite{kirchartzPhotoluminescenceBasedCharacterizationHalide2020}. Consistent with the trends reported by~\cite{othmanAlleviatingNanostructuralPhase2024}, the Cs15 film exhibits both a shift of the PL maximum toward higher photon energies and a higher average QFLS.

The structural interpretation proposed by~\cite{othmanAlleviatingNanostructuralPhase2024} is further supported by the transient-absorption measurements shown in \cref{fig:fig1}c. The ground-state-bleach maximum of the Cs15 film is shifted by approximately \SIrange{20}{30}{\milli\electronvolt}, comparable to the \SIrange{15}{25}{\milli\electronvolt} shift observed in the steady-state PL maximum. The agreement between these two independent optical measurements indicates that the PL shift primarily reflects an increase in the average bandgap of the photoactive Cs15 perovskite phase, rather than arising solely from defect-mediated changes in the emission spectrum\cite{urbanDiscreteDonorAcceptor2023} or from optical reabsorption\cite{fasslRevealingInternalLuminescence2021}. An increase in average bandgap is consistent with a more homogeneous incorporation of Cs into the FA-based perovskite lattice for the Cs15 sample.

By contrast, the Cs0 and Cs30 films exhibit similar low-energy bleach maxima. This suggests that their measured carrier populations are dominated by lower-bandgap grains, despite their different nominal compositions and microstructures. 

Overall, the higher QFLS indicate improved optoelectronic quality of the Cs15 film while the shifted PL and TA bleach maxima indicate a better incorporation of Cs into the perovskite lattice. These observations are consistent with the previous reports of the same samples~\cite{othmanAlleviatingNanostructuralPhase2024}.

\subsection{Fluence-dependent time-resolved photoluminescence}

\begin{figure}[h!]
  \centering
  \includegraphics[width=1\columnwidth]{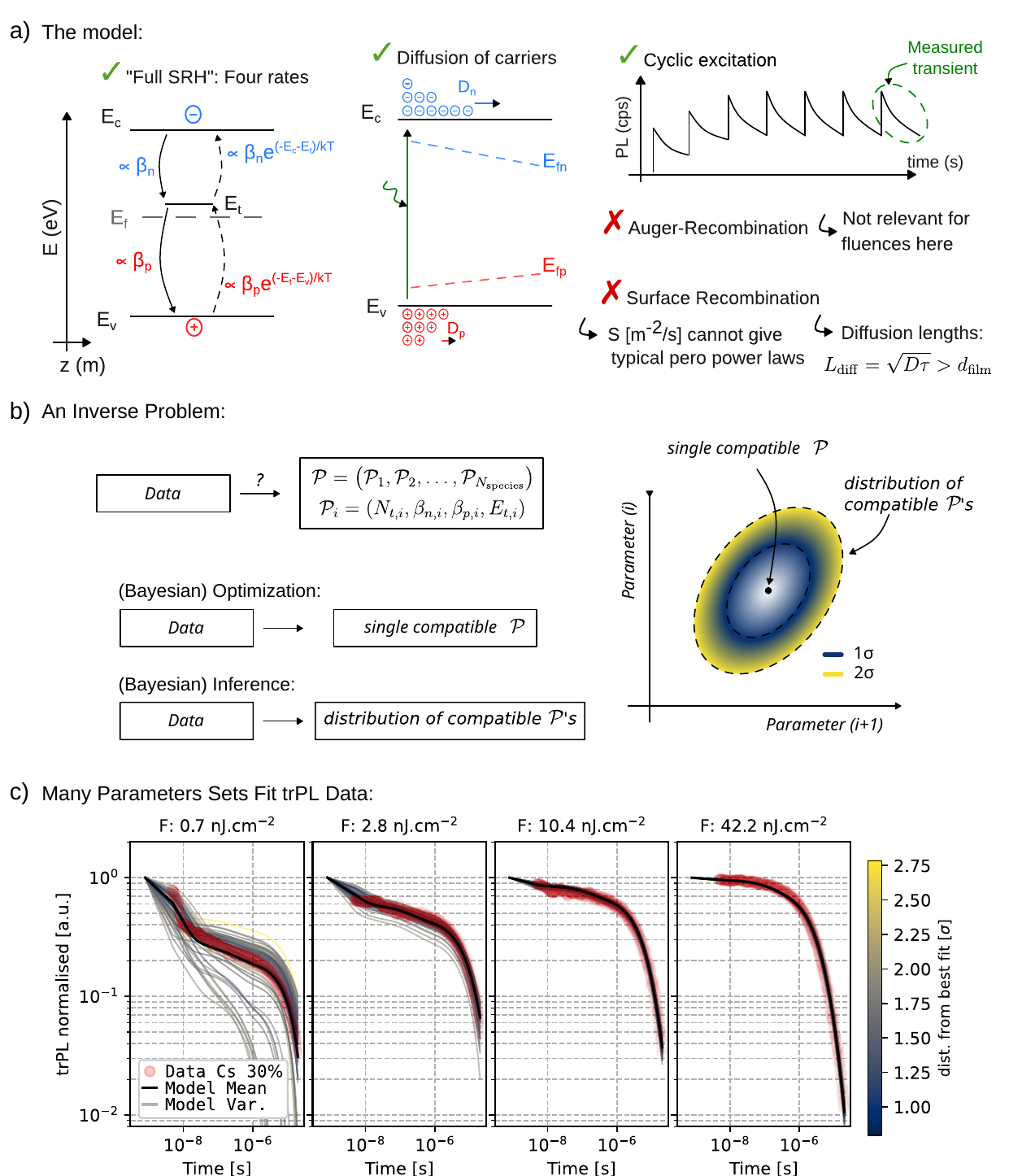}
  \caption{\label{fig:fig2}
Inverse problems can be non-unique. (a) Schematic depiction of model development. (b) Illustration of the inverse problem. Sets of model parameters are compatible with the measured data. (c) For trPL, a wide range of parameter sets can reproduce the measured data. Shown are 100 simulated curves with NRMSE values within 0.01 of the best-fit solution.
}
\end{figure}

We next measure fluence-dependent trPL over an excitation-fluence range of \SIrange{0.7}{45}{\nano\joule\per\centi\meter\squared}, spanning the widest range accessible with our experimental setup. The resulting transients are shown in \cref{fig:fig1}(c). As seen, the Cs0 film exhibits the fastest PL decays across the investigated fluence range, whereas the Cs15 and Cs30 transients become increasingly similar at higher fluences.

In \cref{fig:SI:trPL}b, all three compositions exhibit a pronounced early-time reduction in PL intensity at low excitation fluence. This feature progressively weakens as the fluence is increased. Such behaviour is consistent with trap filling: at low fluence, a substantial fraction of photogenerated carriers can be captured by initially unoccupied shallow states, whereas at higher fluence, these states become increasingly occupied and a smaller fraction of free carriers is captured, reducing their influence on the early-time dynamics~\cite{marunchenkoChargeTrappingDefect2024}.

This strong fluence dependence therefore indicates that the average trap occupation changes substantially across the investigated excitation-fluences. Accurately describing these transients consequently requires a model that explicitly treats carrier capture, emission, and time-dependent trap occupation, motivating the full-SRH framework introduced below.

\subsection{Model framework and assumptions}

As described in the Supplementary Information and in \textcite{simmondsQuantifyingEvolvingDefect2026}, we employ a numerical model based on the full Shockley--Read--Hall formalism~\cite{shockleyStatisticsRecombinationHoles1952}, allowing for an arbitrary number of trapping signatures~\cite{simmonsNonequilibriumSteadyStateStatistics1971}. Electron and hole capture, carrier emission, and the time-dependent trap occupation are treated explicitly. 

As shown in \cref{fig:SI:trPL}, the early-time dynamics contribute substantially to the measured transients and therefore require carrier diffusion to be considered explicitly. Immediately following photoexcitation, photogenerated carriers are distributed according to a Beer--Lambert absorption profile and subsequently diffuse through the film while undergoing radiative recombination and defect-mediated capture and emission.

Auger--Meitner recombination is omitted because no measurable contribution is expected over the excitation-fluence range investigated here, which is representative of carrier densities relevant to photovoltaic operation, see \cref{SI:Auger}.

Surface recombination is not included as a separate boundary condition. Under the commonly used approximation of a constant surface-recombination velocity, $S$, surface recombination reduces to an effective first-order loss term with a carrier-density-independent rate coefficient\cite{lukeAnalysisInteractionLaser1987}, see \cref{SI:Surfaces}. Such a term cannot reproduce the power-law decays frequently observed in perovskite thin films. Given that carrier diffusion lengths in high-quality perovskite films typically exceed the film thickness, we instead describe all nonradiative recombination processes using effective bulk trapping populations treated within the full SRH formalism, allowing power-law dynamics to emerge. The inferred trapping parameters should consequently be interpreted as effective quantities representing the combined influence of bulk and surface-mediated recombination processes present in the experiment.

The long decay times observed at low carrier densities, together with the \SI{5}{\kilo\hertz} repetition rate of the time-correlated single-photon-counting measurements, also require cyclic excitation to be considered. Successive excitation and decay cycles are simulated until a periodic steady state is reached. This procedure ensures that the carrier and trap occupations immediately before each excitation pulse are consistent with the experimental repetition rate, see \cref{SI:Equilibration}.

The full-SRH model therefore provides a physically motivated description of the measured carrier dynamics, but determining its underlying parameters from trPL measurements remains an inverse problem (see \cref{fig:fig2}(b)). In such inverse problems, different combinations of trap densities, capture coefficients, and trap energies may reproduce the same experimental transients with similar accuracy, as illustrated in \cref{fig:fig2}(c), such that a single best-fit solution does not necessarily correspond to a unique physical description of the system. We first assess the model complexity, defined here as the number of trapping signatures included in the model, by systematically varying them and comparing best fit solutions, as detailed in the Supplementary Information \cref{SI:ModelComplexity}.

Across the compositional series, a maximum of three trapping species was required to describe the measured dynamics. For Cs15 and Cs0, increasing the model beyond three trapping species did not yield a significant improvement in NRMSE, whereas for Cs30 no significant improvement was observed beyond two species (\cref{fig:SI:fit_statistics}). We nevertheless use $n_t=3$ for all compositions to maintain a consistent model dimensionality and facilitate direct comparison of the inferred trapping signatures across samples. The resulting best-fit trPL transients are shown in \cref{fig:SI:TuRBO_fits}.

Despite this agreement with experiment, multiple parameter combinations can provide similarly accurate solutions, as shown in \cref{fig:fig2}(c). Individual best-fit parameter values therefore cannot necessarily be interpreted or compared directly. We consequently use MCMC posterior sampling to map the regions of parameter space compatible with the measured data and to quantify correlations between the inferred trapping parameters.

The model, together with the Bayesian-optimisation and posterior-inference routines used in this work, is implemented and available within the OptimPV framework~\cite{vincentm.lecorreOptimPVOptimizationModeling}.

\subsection{Bayesian Inference: Cs30}

We next use Bayesian posterior inference to map the regions of parameter space compatible with the measured trPL transients and to reveal correlations between the inferred parameters. For all analyses presented below, the log-likelihood is defined as

\begin{equation}
\log \mathcal{L} = -\tfrac{1}{2} \left( \frac{\mathrm{NRMSE}}{\sigma} \right)^2,
\end{equation}

with $\sigma=0.005$, corresponding to a characteristic likelihood scale of \SI{0.5}{\percent} in NRMSE. As seen in \cref{fig:fig2}(c), a deviation of 1 sigma from the optimum in a corner plot corresponds to a deviation of NRMSE by $\Delta\sigma=0.005$. The same likelihood definition and $\sigma$'s are used for all compositions, enabling internally consistent comparisons between the sampled parameter distributions.

In the present analysis, posterior sampling is used primarily to identify parameter correlations, degeneracy directions, and recurring regions of parameter space. We therefore focus on the topology and relative overlap of the sampled distributions rather than interpreting their absolute widths as calibrated credible intervals.

\begin{figure}[h!]
  \centering
  \includegraphics[width=1.0\columnwidth]{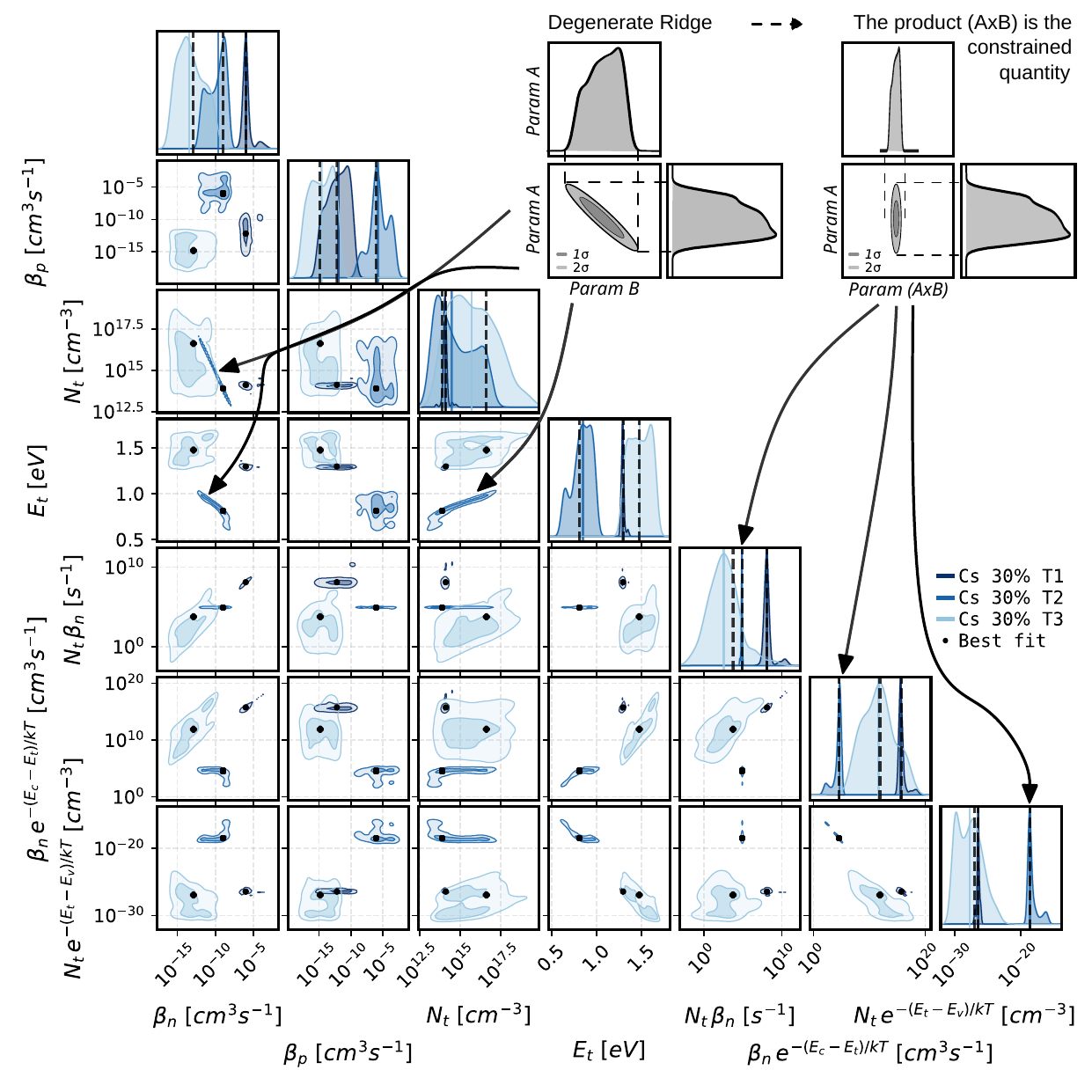}
  \caption{\label{fig:fig3:MCMC-Cs30}Posterior distributions and pairwise parameter correlations for the three effective trapping signatures inferred for the Cs30 film. The complete corner plot, including all fitted model parameters, is provided in \cref{SI:FullCornerPlots}.}
\end{figure}

In \cref{fig:fig3:MCMC-Cs30}, we present the corner plot of Cs30, where three effective trapping signatures are hereafter denoted T1, T2 and T3. The complete corner plots for all samples are provided in \cref{SI:FullCornerPlots}. There, we observe no pronounced correlations between parameters assigned to different trapping signatures, for example between $E_{t,1}$ and $E_{t,2}$. We therefore overlay the posterior distributions of T1, T2 and T3 in each subpanel of \cref{fig:fig3:MCMC-Cs30}. For example, cross-signature projections, such as $E_{t,1}$ versus $N_{t,2}$, are not shown in \cref{fig:fig3:MCMC-Cs30} because they do not show inter-signature correlations. 

As stated in the introduction, it is also important to mention that a trapping signature is not necessarily a single defect site, and that several signatures could eventually arise from single structural defects.

We then turn to the specific trapping signatures observed. First, for T2, we observe a pronounced correlation between $\beta_n$/$N_t$, $\beta_n$/$E_t$, and also $N_t$/$E_t$, respectively. In these three corresponding sub-panels, parameter pairs lying along the ridge of the posterior distribution yield similarly accurate fits, indicating a degeneracy between these quantities. This behaviour follows directly from the SRH formulation for capture and emission of electrons and holes:

\begin{align}
c_{n,i}(t) &= \beta_{n,i}\,n(t)\,\bigl(N_{t,i} - n_{tr,i}(t)\bigr)  \simeq \beta_{n,i}\,N_{t,i}\,n(t)
\label{eq:cn}\\
e_{n,i}(t) &= \beta_{n,i}\,N_c\,\exp\!\left(-\frac{E_c - E_{t,i}}{kT}\right)\,n_{tr,i}(t)
\label{eq:en}\\
c_{p,i}(t) &= \beta_{p,i}\,p(t)\, n_{tr,i}(t)
\label{eq:cp}\\
e_{p,i}(t) &= \beta_{p,i}\,N_v\,\exp\!\left(-\frac{E_{t,i} - E_v}{kT}\right)\,\bigl(N_{t,i} - n_{tr,i}(t)\bigr) \simeq \beta_{p,i}\,N_v\,\exp\!\left(-\frac{E_{t,i} - E_v}{kT}\right)\,N_{t,i}.
\label{eq:ep}
\end{align}

where the approximations above hold when $N_{t,i} \gg n_{tr,i}(t)$, corresponding to weak trap occupation. In these three regimes, the measured dynamics constrain primarily the products $\beta_{n,i}\,N_{t,i}$, $\beta_{n,i}\,\exp\!\left(-\frac{E_c - E_{t,i}}{kT}\right)$, and $N_{t,i}\,\exp\!\left(-\frac{E_{t,i} - E_v}{kT}\right)$, rather than their individual components. This is also evident from the last three corresponding marginal distributions plotted on the bottom right of \cref{fig:fig3:MCMC-Cs30}, which are substantially narrower for $\beta_n\,N_t$,  $\beta_{n,i}\,\exp\!\left(-\frac{E_c - E_{t,i}}{kT}\right)$, and $N_{t,i}\,\exp\!\left(-\frac{E_{t,i} - E_v}{kT}\right)$ than for either $\beta_n$, $E_t$ or $N_t$ individually. Parameter degeneracies of this type are well established in semiconductor defect-parameter extraction~\cite{reinLifetimeSpectroscopyMethod2005}.

In contrast, the next species T1 does not exhibit any pronounced degeneracies, evidenced by more localised posterior probability distributions. For  $\beta_n$/$N_t$, and $N_t$/$E_t$, we attribute this reduced degeneracy to significant trap filling. When $n_{tr,i}(t)$ becomes comparable to $N_{t,i}$, the approximation in \cref{eq:cn} no longer holds, and the capture rate depends explicitly on the instantaneous number of unoccupied states, $N_{t,i}-n_{tr,i}(t)$. Significant trap occupation consequently lifts the $\beta_n$--$N_t$ degeneracy, allowing the capture coefficient and trap density to be more independently constrained. While defect occupation is known to influence semiconductor lifetime measurements~\cite{mcintoshGeneralizedProcedureDetermine2008}, our analysis shows that it also determines whether the capture-coefficient--trap-density degeneracy persists.

For T1, significant trap filling also makes the dynamics sensitive to experimental conditions that determine the trap occupation, particularly excitation fluence and repetition rate. Changes in either quantity alter the time-dependent trap population $n_{tr,i}(t)$, further emphasizing the importance of accounting for cyclic excitation when modelling trPL measurements in regimes where trap populations do not fully relax between successive excitation pulses~\cite{trimplChargeCarrierTrappingRadiative2020a,vincentm.lecorreOptimPVOptimizationModeling}.

Finally, we observe that T1 is a trapping signature that gives rise to the characteristic $1/n$ dependence in differential-decay versus mean-carrier-density plots (see \cref{SI:1overN}), which has been shown to be the limiting case for power-law behaviours observed in experimental perovskite trPL~\cite{yuanUnderstandingPowerLawPhotoluminescence2024}.

\subsection{Posterior overlap identifies common trapping signatures across samples}

\begin{figure}[h!]
  \centering
  \includegraphics[width=1\columnwidth]{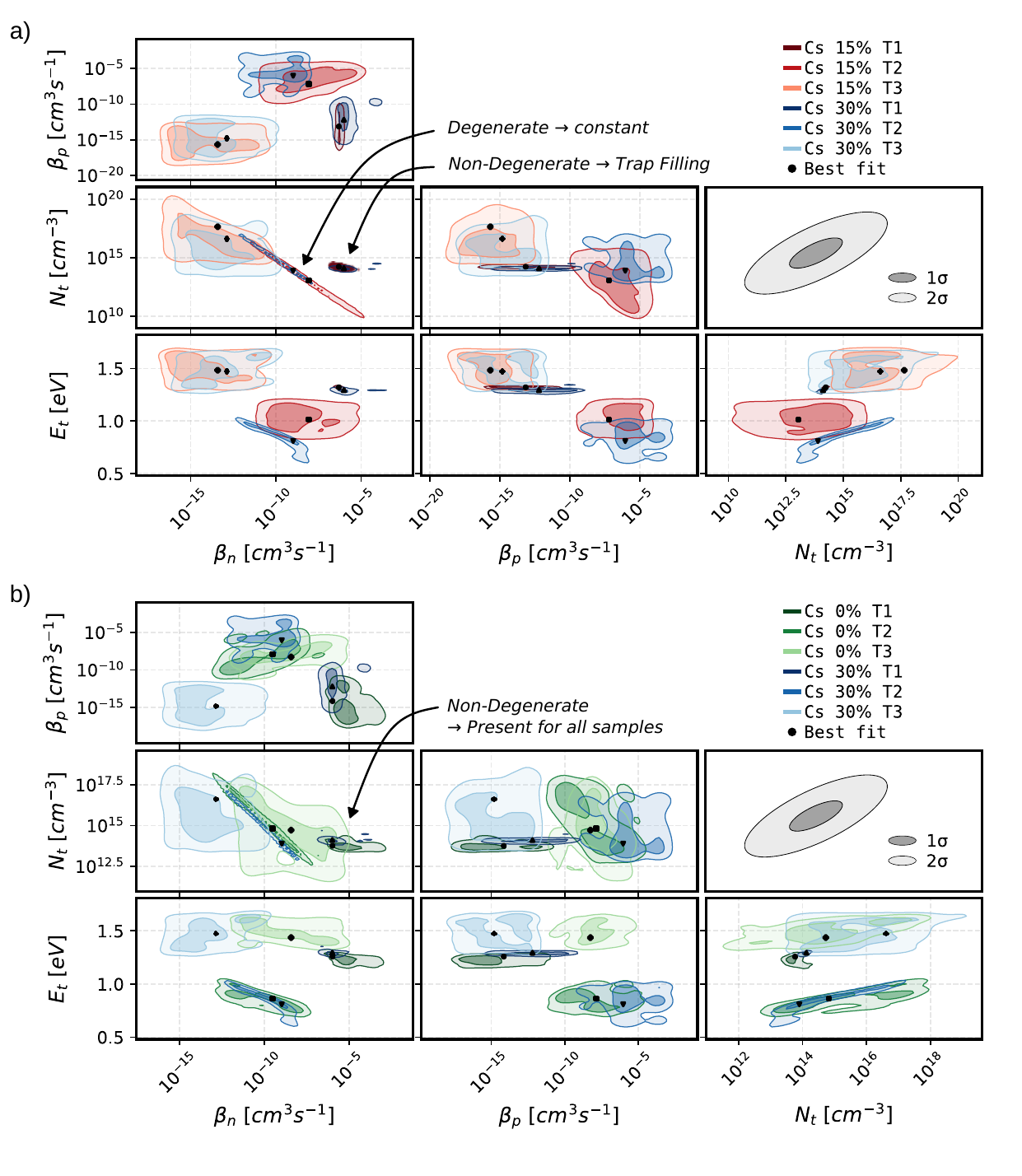}
  \caption{\label{fig:fig4} Posterior distributions and pairwise parameter correlations for the effective trapping signatures inferred for (a) Cs15 and Cs30 films and (b) Cs0 and Cs30 films. (a) Both T1 and both T3 posterior distributions exhibit substantial overlap between the two compositions. (b) Both T1 and both T2 distributions overlap between the two compositions. In contrast to the previous sample pair, the T3 distributions do not overlap.}
\end{figure}

We next extend the posterior analysis across the FA$_{1-x}$Cs$_x$PbI$_3$ compositional series. In \cref{fig:fig4}(a), the posterior distributions obtained for Cs15 are shown together with the three trapping signatures identified for Cs30 (\cref{fig:fig3:MCMC-Cs30}). In contrast to the individual best-fit parameter values obtained by Bayesian optimisation (black dots), the posterior distributions reveal extended regions of parameter space that are compatible with the measurements. Overlap between these regions therefore enables trapping signatures to be identified and compared directly across the two compositions.

As shown in \cref{fig:fig4}(a) of the $\beta_n$ and $N_t$ panel, both T1 and both T3 posterior distributions exhibit substantial overlap between the Cs15 and Cs30 samples. Notably, as previously observed for Cs30, T1 again exhibits little to no degeneracy for Cs15. The similarity in both posterior location and correlation structure suggests that T1 and T3 represent common effective trapping signatures present in both compositions. This correspondence is also observed across the remaining posterior projections.

By contrast, a third trapping signature, T2, occupies a shifted region of parameter space for Cs15, particularly in the $E_t$-$N_t$ and the $E_t$-$\beta_n$ projections. For this signature, we also observe a pronounced $\beta_n\,N_t$ degeneracy ridge with a shift towards higher $\beta_n$ values or lower $N_t$ values. In contrast to Cs30, however, the $\beta_n$--$E_t$ and $N_t$--$E_t$ degeneracies are no longer observed for Cs15, while the posterior is shifted toward higher trap energies.

The Cs30 composition has previously been reported to contain segregated $\delta$-CsPbI$_3$ secondary phases, whereas the intermediate Cs15 composition exhibits a more homogeneous Cs-containing perovskite microstructure\cite{othmanAlleviatingNanostructuralPhase2024}. Therefore, the presence of T1 and T3 in both compositions suggests that these signatures may originate from homogeneous Cs-containing perovskite grains, which are likely present in both samples. By contrast, we observe shifts in the T2 signature towards more efficient electron capture (increase $\beta_n$) and increase in trap energy, which may arise from structural differences between the samples, such as the better incorporation of Cs observed in \cref{fig:fig1}. Overall, these observations highlight the strength of the methodology described above in consistently identifying material properties, enabling their correlation with independently observed structural features.

Finally, we compare the posterior distributions of Cs30 (blue) and Cs0 (green) in \cref{fig:fig4}(b). As in the previous comparison panel (a), the T1 distributions occupy similar regions in the parameter space, with a shift for the Cs0 to slightly higher $\beta_n$. These corresponding non-degenerate trapping signatures, which we have associated with significant trap filling, therefore are present across all compositions, consistent with the observation of pronounced fluence-dependent trap-filling behaviour in all the trPL measurements of \cref{fig:SI:trPL}.

In contrast to the previous comparison, the T2 distributions also exhibit substantial overlap between Cs30 and Cs0. As discussed above, this trapping signature shows pronounced correlations between $\beta_n$ and $N_t$, $\beta_n$ and $E_t$, and $N_t$ and $E_t$. Because these distributions arise in the Cs0 that contains no Cs and the Cs30 that contains segregated grains of $\delta$-CsPbI$_3$, we tentatively attribute this signature to Cs-poor grains, which are present in both samples, according to the TA measurements of \cref{fig:fig1}, where the photoactive phases show similar bandgap.

Additionally, the T3 distributions exhibit little overlap between the two compositions and are characterised by broader posterior distributions without a pronounced correlation structure. The location of the Cs0 T3 distribution in a distinct region of parameter space correlates with the high density of stacking faults, or ${111}_{\mathrm{C}}$ planar defects, providing an early indication of the defect properties associated with these structures.

To summarize the cross-composition comparison, we identify four common or composition-specific trapping signatures. First, the T1 distributions of all three compositions occupy similar regions in parameter space and are therefore grouped into a common trapping signature, denoted $T\beta$. $T\beta$ exhibits little to no $\beta_n$–$N_t$ parameter correlation ridge. As discussed above, the absence of this product degeneracy is consistent with significant trap filling in this regime.

Second, the T2 distributions of Cs0 and Cs30 largely overlap, whereas Cs15 exhibits a shifted overlap toward higher $\beta_n$ and $E_t$. We denote this signature as $T\alpha$. In contrast to $T\beta$, $T\alpha$ exhibits pronounced parameter degeneracies, most consistently between $\beta_n$ and $N_t$, with additional $\beta_n$–$E_t$ and $N_t$–$E_t$ correlations observed for Cs0 and Cs30.

Third, the T3 distributions of Cs30 and Cs15 overlap, defining a further common trapping signature, denoted $T\gamma$. These distributions exhibit no pronounced parameter correlations.

Finally, T3 of Cs0 forms a distinct, composition-specific trapping signature that does not occupy a region in parameter space observed in the other samples. We name T3 for Cs0 $T\epsilon$.

Overall, this comparison illustrates the value of posterior distributions for identifying recurring trapping signatures across different samples. If only the individual best-fit solutions were compared, as indicated by the black markers in \cref{fig:fig4}, the correspondence between trapping species across compositions would be difficult to establish. The posterior distributions instead reveal regions of parameter space compatible with experimental data, as well as their associated parameter correlations, enabling more robust cross-sample comparison.

\begin{table}[h!]
\centering
\caption{\label{tab:trapping-signatures}Summary of the common and composition-specific effective trapping signatures identified across the FA$_{1-x}$Cs$_x$PbI$_3$ series.}
\begin{tabular}{l|l|l|l|l}
\hline
\textbf{Signature} & \textbf{T$_{i}$} & \textbf{Sample} & \textbf{Parameter degen.} & \textbf{Tentative assignment} \\
\hline
$T\beta$   & T1 & Cs0, Cs15, Cs30 & No degen. & Significant trap filling \\
$T\alpha$  & T2  & Cs0, Cs15, Cs30 & $\beta_n$--$N_t$ & \\
$T\gamma$  & T3 & Cs15, Cs30 & No degen. & Homogeneous Cs-containing grains \\
$T\epsilon$ & T3 (Cs0) & Cs0 only & No degen. & Composition-specific \\

\end{tabular}
\end{table}

\subsection{Trapping parameters identify \(T\beta\) as associated with trap filling and \(T\epsilon\) as detrimental to optoelectronic quality.}

The analysis above identifies four effective trapping signatures across the measured samples. Their corresponding parameter distributions are summarized in \cref{fig:SI:MCMC-Boxplot}. Consistent with the posterior-overlap analysis, the parameter distributions associated with each signature show clear clustering in the boxplot representation.

\begin{figure}[h!]
\centering
\includegraphics[width=1\textwidth]{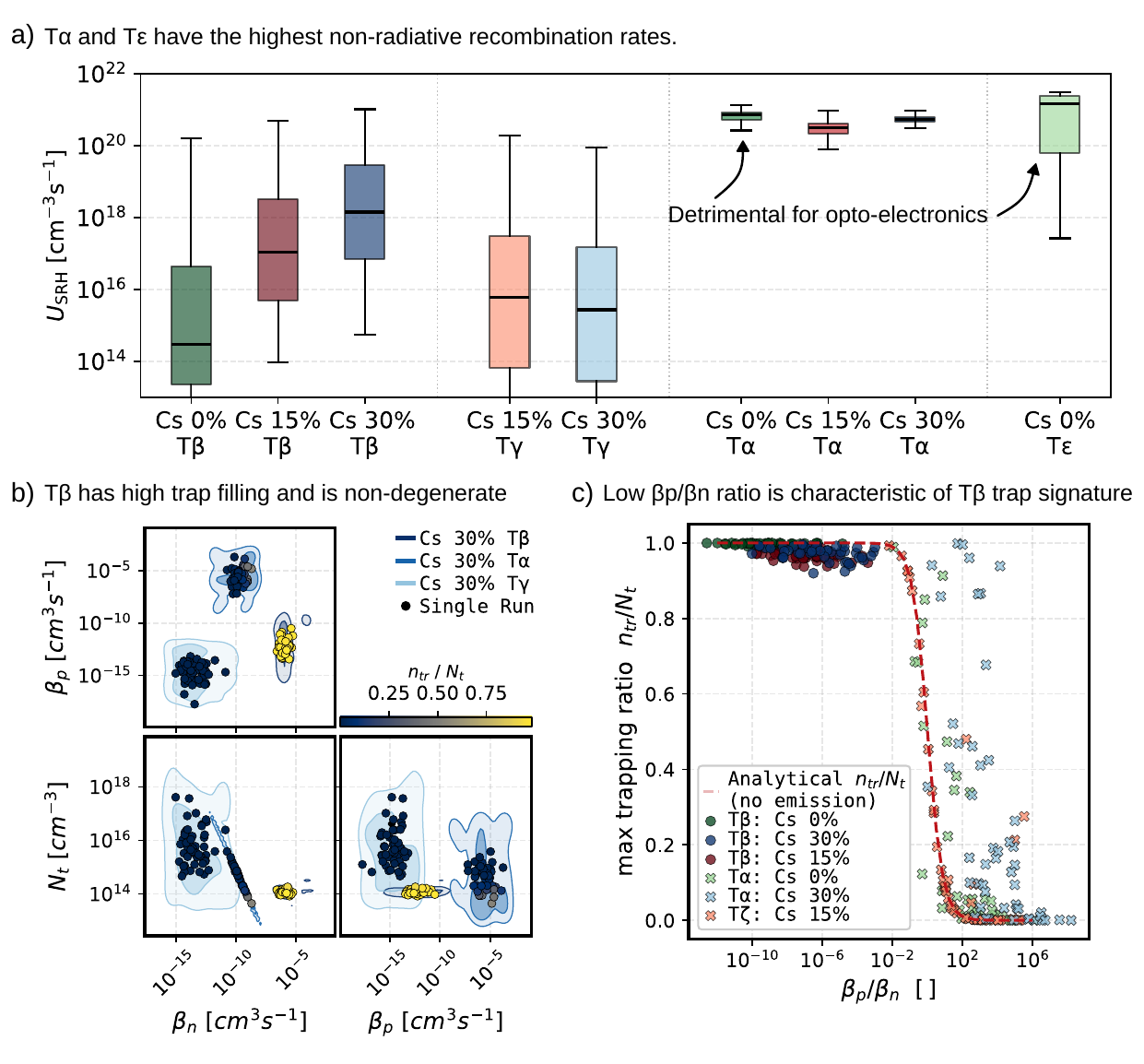}
\caption{\label{fig:fig5}
(a) Steady-state SRH recombination rates calculated from the inferred parameter distributions for the four effective trapping signatures.
(b) Maximum trap filling ratio reached during the transient for each parameter set. The parameter regions associated with \(T\beta\) exhibit substantial trap filling.
(c) \(\beta_p/\beta_n\) as a function of the maximum trap filling ratio. Very low \(\beta_p/\beta_n\) ratios are characteristic of the \(T\beta\) signatures observed across all perovskite compositions studied here.}
\end{figure}

First, the steady-state SRH recombination rates associated with these parameter distributions are shown in the final panel of \cref{fig:fig5}. The procedure used to propagate the posterior samples to steady-state conditions is described in \cref{SI:Usrh}. Notably, the $T\epsilon$ and $T\alpha$ signatures exhibit high non-radiative recombination rates. For Cs30, the lower QFLS observed in \cref{fig:fig1} can therefore be attributed to a higher trap density of the $T\alpha$ signature, as compared to Cs15. In contrast, the lower QFLS in Cs0 is associated with the emergence of a new $T\epsilon$ signature, which, somewhat counterintuitively, is characterized by a substantially shallower trap energy.

Here, the \(T\alpha\) signature emerges as particularly important for device design, as it constitutes a major non-radiative recombination channel and is present in all samples investigated. This signature also exhibits strong parameter degeneracies, highlighting the importance of explicitly accounting for parameter correlations and uncertainty in the inverse problem to obtain a reliable physical interpretation.

Finally, all samples investigated exhibit the \(T\beta\) signature. In \cref{fig:fig5}(b), we further show that trapping states within this region of parameter space are characterized by pronounced trap filling, which arises from a very low \(\beta_p/\beta_n\) ratio.

More generally, these results indicate that a distinctive feature of perovskite optoelectronics is the presence of trapping signatures with strongly asymmetric capture coefficients, i.e., very low \(\beta_p/\beta_n\) ratios. This asymmetry results in substantial trap filling during transient measurements, highly affecting the final measured decays.

Overall, these effects may also explain the strong sensitivity of such measurements to experimental conditions, such as excitation fluence and repetition rate, both of which directly influence trap occupancy. Moreover, sufficiently long-lived trap occupation may provide the timescale required for chemical reactions to occur while a trap remains filled. This connection could provide a route toward identifying which trapping signatures are associated with defects that evolve under operational stress~\cite{simmondsQuantifyingEvolvingDefect2026}.

\newpage

\section{Conclusion}

By combining full Shockley--Read--Hall modelling with posterior sampling, we show that multiple defect-parameter combinations can reproduce the measured fluence-dependent trPL and that the identifiability of these parameters depends strongly on the underlying trapping regime. This analysis demonstrates why individual best-fit parameter values can lead to misleading physical interpretations when parameters are weakly constrained or strongly correlated. 

First, we identify pronounced correlations between \(\beta_n\) and \(N_t\) for several trapping species. Under conditions of low trap occupation, the measured dynamics predominantly constrain the product \(\beta_n\,N_t\), rather than its individual components. By contrast, significant trap filling lifts this degeneracy by making the capture dynamics explicitly sensitive to the number of unoccupied trapping states. Importantly, the resulting non-degenerate \(T\beta\) signature, characterized by a very low \(\beta_p/\beta_n\) ratio, is observed across all three FA$_{1-x}$Cs$_x$PbI$_3$ compositions, suggesting that this trapping regime is a shared feature of the investigated perovskite films. This signature also produces the characteristic power law decays observed and described in~\cite{yuanShallowDefectsVariable2024,yuanUnderstandingPowerLawPhotoluminescence2024}.

Second, by comparing the sampled parameter distributions across the compositional series, we identify four shared or composition-specific effective trapping signatures. Notably, \(T\alpha\) emerges as an important limitation to optoelectronic device performance, as it constitutes a major non-radiative recombination channel in all measured compositions. That said, it features pronounced parameter degeneracies, highlighting the importance of accounting for posterior correlations and uncertainty when interpreting trapping parameters.

In contrast, the shallowest signature, \(T\epsilon\), is observed only in pure FAPbI$_3$ and produces the largest steady-state non-radiative recombination rate. This signature is consistent with a high density of twinning-related stacking faults or $\{111\}_{\mathrm{C}}$ planar defects previously reported for Cs-free films, although a direct microscopic assignment cannot be made from trPL alone.

More generally, our results show that individual best-fit solutions are insufficient for constructing a physically meaningful defect landscape from trPL measurements. Posterior correlation structures and comparisons between fit-compatible parameter distributions provide a more robust basis for identifying and comparing trapping signatures contained in the experimental data.

The identified \(\beta_n\)--\(N_t\) degeneracies also provide guidance for the design of future measurements and mirror the capture-cross-section--defect-density degeneracies known from silicon defect spectroscopy. In silicon, combining injection- and temperature-dependent measurements provides additional sensitivity to correlated defect properties~\cite{reinLifetimeSpectroscopyMethod2005}. Similarly, introducing complementary experimental observables may help independently constrain capture coefficients, trap densities, and trap energies in perovskites.

Finally, the identified parameter degeneracies originate directly from the structure of the Shockley--Read--Hall equations and are therefore not specific to the FA$_{1-x}$Cs$_x$PbI$_3$ system investigated here. Our results further show that sufficient trap occupation determines whether these degeneracies persist, establishing trap filling as a key determinant of defect-parameter identifiability.  The approach therefore provides a route to identifying and comparing trapping signatures across perovskites using room-temperature measurements and, more broadly, across other semiconductor systems governed by Shockley--Read--Hall recombination.

\section{Methods}

\subsection{FACs compositional series}

\noindent The samples investigated here were newly fabricated by the authors of \textcite{othmanAlleviatingNanostructuralPhase2024}.

\textit{Ink preparation}: The perovskite composition used is 
\ce{Cs_{x}FA_{1-x}PbI_3}. 1 M solutions of both neat \ce{FAPbI_3} and \ce{CsPbI_3} films were prepared in mixed solvents of DMF/DMSO (4:1 in v:v) and then mixed by volume to get the desired nominal molar concentrations. Both precursor solutions contain a \SI{10}{\mol\percent} excess \ce{PbI_2}. 

\textit{Thin film preparation}: A perovskite film with $\sim\SI{350}{\nano\meter}$ thickness was then deposited on cleaned glass substrates by spin coating \SI{100}{\micro\liter} of the 1 M precursor solution, first at \SI{1000}{\rpm} (\SI{200}{\rpm\per\second}) for \SI{10}{\second}, followed by \SI{5000}{\rpm} (\SI{800}{\rpm\per\second}) for \SI{35}{\second}. For the antisolvent quenching, \SI{5}{\second} before the end of the spin coating, \SI{300}{\micro\liter} of EA was quickly dropped onto the perovskite surface. The films were then annealed at \SI{150}{\degreeCelsius} for \SI{30}{\minute}. All the film and device fabrication processes were done inside a glovebox filled with \ce{N2} atmosphere. 

\subsection{Steady state PL measurements}
For the steady-state PL measurements of \cref{fig:fig1}, a \SI{532}{\nano\meter} laser is directed via a parabolic mirror (Thorlabs, MPD019-P01) onto the sample. The emitted photoluminescence (PL) is collected using two plano-convex lenses and coupled into an optical fiber connected to a spectrometer (QE Pro, Ocean Insight). The signal was recorded using software from Quantum Yield Berlin (QYB).

The laser (Insaneware) provided an excitation fluence of approximately \SI{0.5}{sun} (for \SI{1.68}{\electronvolt} bandgap), illuminating an area of \SI{0.35}{\centi\meter\squared}. Illumination was performed from the substrate side. For determining the quasi-Fermi levels, the high-energy tail fit method is used~\cite{kirchartzPhotoluminescenceBasedCharacterizationHalide2020}.

\subsection{Time resolved photoluminescence measurements (trPL)}

\noindent For measurement of trPL we used time correlated single photon counting (TCSPC). The setup employs an ``80:20'' transmission:reflection beam splitter to separate the excitation and detection paths. Excitation is provided from the substrate side by a \SI{705}{\nano\meter} diode laser (IB-705-B laser head with Taiko driver, PicoQuant) with a pulse duration of $\sim \SI{100}{\pico\second}$ and a repetition rate of \SI{5}{\kilo\hertz}. The laser beam is passed through a cleanup filter (FF01-700/13--25, Semrock), and its power is adjusted using a linear-gradient neutral density filter to values of $\sim \SIrange{0.002}{0.126}{\micro\watt}$ on the sample.

An off-axis parabolic mirror with a focal length of \SI{5}{\centi\meter} is used for both focusing and PL collection. The laser spot size was measured to be $\sim \SI{250}{\micro\meter}$ in diameter using a CCD camera and a gridded substrate. Photoluminescence (PL) is detected using a silicon single-photon avalanche diode (Laser Components COUNT50), with the signal spectrally filtered by a \SI{715}{\nano\meter} long-pass filter (FF01-715/LP-25, Semrock). The PL count rate and decay histograms are recorded using a TimeHarp~260 Nano TCSPC module (PicoQuant).

The relatively low repetition rate is chosen to enable the observation of long-lived, tailing power-law-like decays characteristic of perovskite thin films. For power-dependent trPL measurements, a representative code snippet for automated acquisition is provided in~\cite{simmondsScriptsUsedThesis2026}. 

\subsection{Transient absorption measurements}

\noindent Transient absorption (TA) spectroscopy was performed using a custom-built setup in standard transmission geometry with near-orthogonal incidence of pump and probe light at the sample surface. Laser pulses were generated and amplified by a commercial Ytterbium-fiber laser system (Light Conversion, CARBIDE-CB3-40W) and amplified at a repetition rate of 20 kHz. Fundamental pulses of 190 fs duration at 1030 nm central wavelength were split to provide the pump and probe pulses, respectively. Pump pulses at 700nm wavelength were created using an optical parametric amplifier (OPA, Light Conversion, ORPHEUS-HP). The remainder of the fundamental pulse energy was utilized to drive a supercontinuum generation process in a 4mm thick YAG window. The probe light thus generated spanned a wavelength range from \SIrange{480}{950}{\nano\meter}, which was focused into the sample after removing the strong driver field with appropriate shortpass filters and enforcing vertical polarization with a wire-grid polarizer (Thorlabs, WP25M-UB1). Pump pulse polarization and intensity was controlled by a combination of half-wave-plate and polarizer. All measurements were performed under magic-angle polarization conditions. The pump fluence was kept at \SI{13}{\micro\joule\per\centi\meter\squared}.
After passing through the sample, probe light was collected by a multimode optical fiber and detected with a fast CCD array (Entwicklungsbüro Stresing, HSVIS) attached to a Czerny-Turner spectrograph (Andor Technology, Shamrock 303).
Transient spectra were recorded up to a pump-probe delay of 1000 ps using a mechanical delay stage. A semi-logarithmic delay sampling was utilized, providing a dense linear sampling during the ultrafast period, and logarithmic sampling afterwards. Time-zero dispersion was modelled using a second order polynomial on the photon-energy scale using the data points generated when the transient signal reaches 50\si{\percent} of its maximum absolute amplitude, and the data were thus corrected. The spectra in \cref{fig:fig1} are plotted at a time delay of \SI{1}{\pico\second}

\subsection{Bayesian Optimisation for global fitting}

\noindent For trPL fitting, the data are first interpolated to reduce the number of points per decay, using time points that are equally spaced on a logarithmic time scale. For the implementation of Bayesian optimisation, we use the framework introduced in~\cite{vincentm.lecorreOptimPVOptimizationModeling}. For the assessment of the goodness of fit, we employ the normalised root-mean-square error (NRMSE) evaluated on normalised and logarithmically transformed data. Since trPL decays span several orders of magnitude, a linear NRMSE would be dominated by the early-time peak and would largely neglect the decay tail. The Bayesian optimisation algorithm used is Trust Region Bayesian Optimisation (TuRBO)~\cite{erikssonScalableGlobalOptimization2019}, which is implemented via the Ax/BoTorch platform~\cite{bakshyAEDomainagnosticPlatform2018}. The optimisation is initialised using a Sobol sequence with 20 initial evaluations to ensure broad coverage of the parameter space. Subsequent optimisation is performed in batches of 4 parallel evaluations over 300--600 iterations, resulting in a total of up to 2420 model evaluations. A GP surrogate model with a Matern 5/2 kernel and automatic relevance determination (ARD) is employed to model the local objective functions. The acquisition strategy is based on the batch noisy expected improvement (qLogNEI), which is well suited for noisy objectives and large dynamic ranges, as encountered in trPL data. The TuRBO trust-region mechanism is controlled via a failure tolerance parameter of 4, determining when the trust region is contracted following unsuccessful optimisation steps.

The trPL error is defined as
\begin{align}
\tilde{PL}_i^{\mathrm{data}} &= \log_{10}\left(\frac{PL_i^{\mathrm{data}}}{PL_{\max}^{\mathrm{data}}}\right),
\qquad
\tilde{PL}_i^{\mathrm{sim}} = \log_{10}\left(\frac{PL_i^{\mathrm{sim}}}{PL_{\max}^{\mathrm{sim}}}\right)
\\[6pt]
\mathrm{NRMSE}_{\mathrm{PL}} &= \frac{\sqrt{\frac{1}{N}\sum_i\left(\tilde{PL}_i^{\mathrm{data}}-\tilde{PL}_i^{\mathrm{sim}}\right)^2}}{\max_i\left(\tilde{PL}_i^{\mathrm{data}},\tilde{PL}_i^{\mathrm{sim}}\right)-\min_i\left(\tilde{PL}_i^{\mathrm{data}},\tilde{PL}_i^{\mathrm{sim}}\right)}.\label{ch3:eq:NRMSEPL}
\end{align}

Two additional constraints are imposed during the fitting procedure:
\begin{itemize}
\item For all trap species, values of the single-exponential decay time \[\tau_i = \frac{1}{N_{\mathrm{t},i}\, \beta_{\mathrm{n},i}}\] are restricted to $\tau_i < \SI{10}{\micro\second}$. This constraint excludes trap species that are too slow to influence the experimentally observed dynamics.

\item After each model evaluation, a maximal penalty is applied to the objective function if the initial fluence-dependent slope $s$ of the trPL count maxima ($t = 0$) is lower than 1.7. This constraint ensures that the simulated trPL decays follow the expected fluence dependence for intrinsic semiconductors, where $s \approx 2$.
\end{itemize}

\subsection{Markov Chain Monte Carlo posterior sampling}

\noindent While Bayesian optimisation efficiently locates the best-fit parameter set, it does not characterise the uncertainty or correlations between fitted parameters. To quantify the posterior probability distribution of the rate-equation model parameters or drift-diffusion parameters, we perform Markov Chain Monte Carlo (MCMC) sampling using the affine-invariant ensemble sampler implemented in the \texttt{emcee} package~\cite{foreman-mackeyEmceeMCMCHammer2013}. 

The log-likelihood is defined as\[\log \mathcal{L} = -\tfrac{1}{2} \left( \frac{\mathrm{NRMSE}}{\sigma} \right)^2,\] where the same normalised, log-transformed NRMSE used in the Bayesian optimisation step is employed, and $\sigma$ represents an effective noise floor reflecting the typical fit accuracy achievable by the model. A fixed value of $\sigma = 0.005$ is used across all MCMC runs to ensure that the resulting posteriors are directly comparable. This likelihood formulation corresponds to assuming approximately Gaussian-distributed residuals in logarithmic space.

Walkers are initialised in a tight Gaussian cluster around the TuRBO best-fit parameters, ensuring the exploration of the minima found by TuRBO. 
For each sampling run, we use $N_{\mathrm{walkers}} \geq 2 N_{\mathrm{dim}}$ walkers and perform $N_{\mathrm{steps}}$ iterations following a burn-in phase. The total number of steps is chosen such that the integrated autocorrelation time $\tau$ satisfies $N_{\mathrm{steps}} \gtrsim 50\,\tau$ whenever feasible.

Convergence is assessed by comparing the parameter medians between the last quarter and second to last quarter of each chain; only the converged portion (last half) is retained for posterior analysis. The resulting samples are visualised as corner plots showing the marginalised one- and two-dimensional posterior distributions, from which parameter distributions and correlation structures are analysed. All code can be found in~\cite{simmondsScriptsUsedThesis2026}.

\subsection{Steady state propagation of parameters}

For estimation of the recombination properties associated with the inferred trapping parameters, we assume a constant photon flux of \(N_0 = \SI{1.5e17}{\per\centi\meter\squared\per\second}\), corresponding approximately to a 1-Sun photon flux for a bandgap of \SI{1.55}{\electronvolt}. For each posterior parameter set, the steady-state carrier densities are determined and the corresponding defect-specific Shockley--Read--Hall recombination rate \(U^{\mathrm{SS}}_{\mathrm{SRH},i}\) is evaluated for each defect class. This allows us to assess which defects contribute most strongly to non-radiative recombination under solar-cell operating conditions.

For defect class \(i\), the SRH recombination rate is calculated according to~\cite{shockleyStatisticsRecombinationHoles1952}
\begin{align}
U_{\mathrm{SRH},i} =
N_{t,i},
\frac{\beta_{n,i}\beta_{p,i}\left(np-n_i^2\right)}
{\beta_{n,i}\left(n+n_{1,i}\right)+
\beta_{p,i}\left(p+p_{1,i}\right)}.
\label{eq:SS-SRH}
\end{align}

The steady-state carrier densities \(n\) and \(p\) are obtained numerically by simultaneously solving the charge-neutrality condition for the occupied electron traps,
\begin{equation}
p-n-\sum_{i=1}^{n_{\mathrm{traps}}} n_{\mathrm{tr},i}=0,
\label{eq:eqUSRH_chargeNeutr}
\end{equation}
and the total generation--recombination balance,
\begin{equation}
G-R_{\mathrm{rad}}-\sum_{i=1}^{n_{\mathrm{traps}}}U_{\mathrm{SRH},i}=0,
\label{eq:eqUSRH_Gen}
\end{equation}
where the radiative recombination rate is
\begin{equation}
R_{\mathrm{rad}} = k_{\mathrm{rad}}np.
\end{equation}

Thus, although \(U^{\mathrm{SS}}_{\mathrm{SRH},i}\) denotes only the non-radiative SRH contribution of an individual defect class, its steady-state value is evaluated using carrier densities that result from competition between radiative recombination and the SRH recombination of all defect classes.

\newpage
\appendix
\setcounter{secnumdepth}{1}
\setcounter{figure}{0}

\renewcommand{\thesection}{S\arabic{section}}
\renewcommand{\thefigure}{S\arabic{figure}}
\renewcommand{\theHfigure}{S\arabic{figure}}

\section{Previous TEM measurements}
\label{SI:TEM}
\begin{figure}[h!]
  \centering
  \includegraphics[width=0.9\columnwidth]{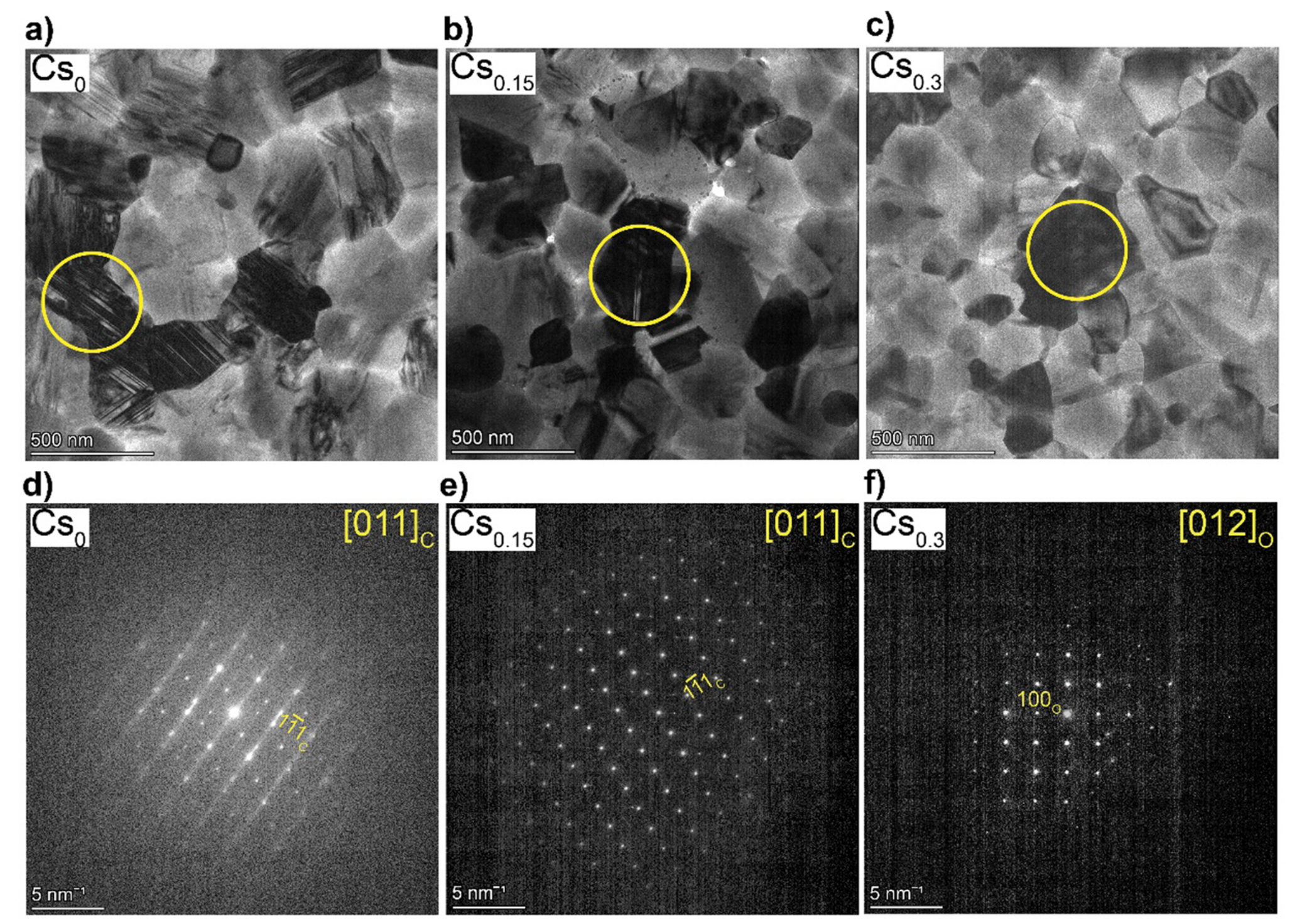}
  \caption{\label{fig:SI:cs-doping}Previously reported microstructural evolution across the FA$_{1-x}$Cs$_x$PbI$_3$ compositional series, reproduced from \textcite{othmanAlleviatingNanostructuralPhase2024}. Cs-free films exhibit a high density of twinning-related stacking faults, the intermediate-Cs composition minimizes the observed structural defects, and the high-Cs composition contains Cs-rich nonperovskite secondary phases.}
\end{figure}

\section{Early Time Dynamics in trPL Decays a Typical Indication of Trap Filling.}

\begin{figure}[h!]
  \centering
  \includegraphics[width=1\columnwidth]{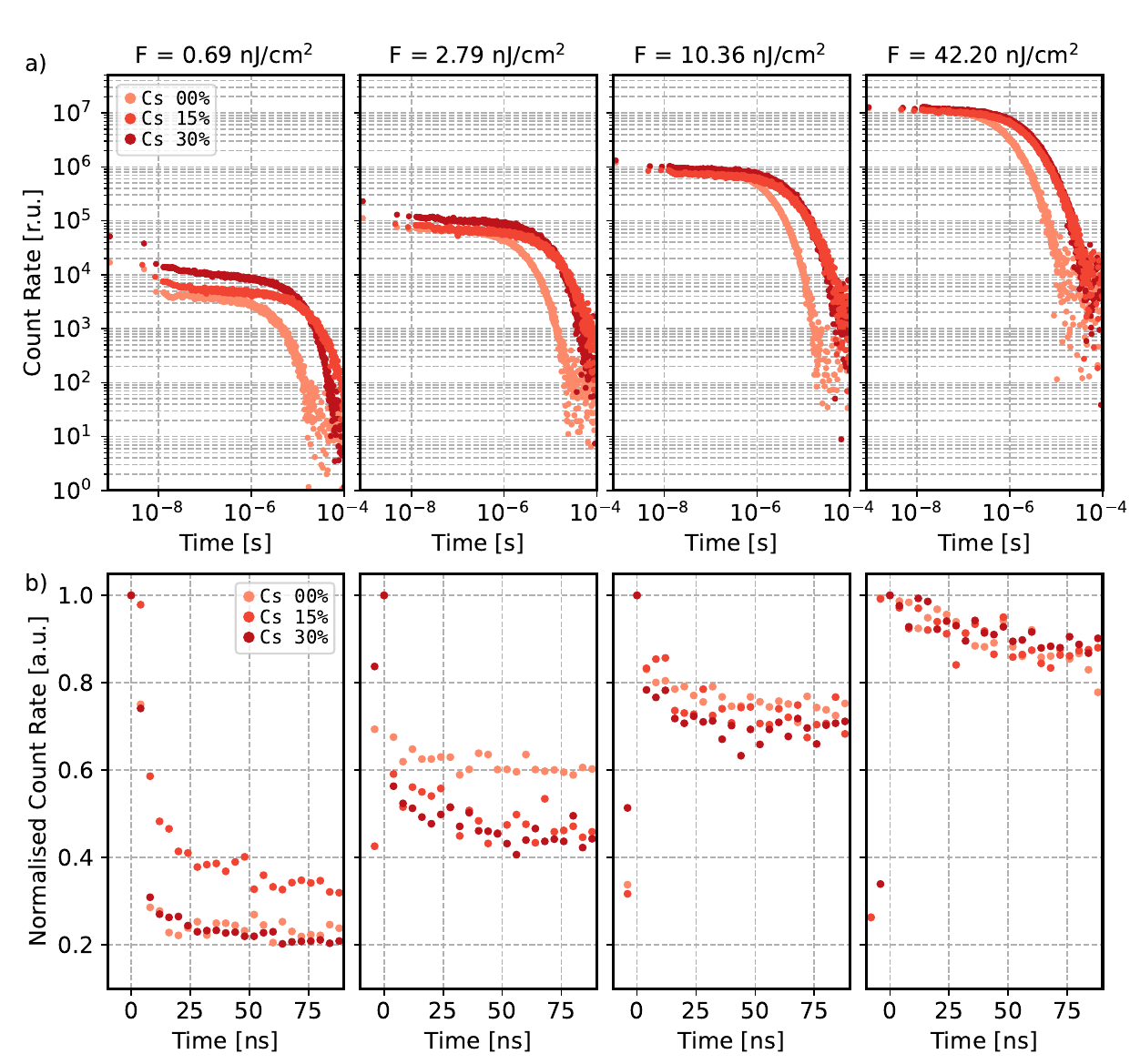}
  \caption{\label{fig:SI:trPL}
Fluence-dependent time-resolved photoluminescence measurements of the Cs0, Cs15, and
Cs30. (a) Plotted on a log--log scale. (b) The early-time PL drop exhibits a fluence dependence. At higher fluences, this drop is no longer observed, consistent with trap filling.
}
\end{figure}

\newpage

\section{Model Rate Equations}\label{SI:ModelRateEq}
\noindent For trPL modelling, the system of differential equations \cref{SI:eq:SRHrateEq1,SI:eq:SRHrateEq2,SI:eq:SRHrateEq3}, which follows the SRH description\cite{shockleyStatisticsRecombinationHoles1952,simmonsNonequilibriumSteadyStateStatistics1971}, are solved using the \texttt{solve\_ivp} routine from SciPy~\cite{virtanenSciPy10Fundamental2020}. 

Two numerical integration methods are employed: LSODA~\cite{petzoldAutomaticSelectionMethods1983}, which automatically switches between stiff and non-stiff regimes, and a backward differentiation formula (BDF) method~\cite{shampineMATLABODESuite1997}, which is particularly suited for stiff systems. An analytical Jacobian is implemented and verified, enabling faster evaluation of local derivatives of the rate equations, significantly improving computational performance. Despite this, the model exhibits highly stiff regimes spanning multiple timescales, which can hinder convergence of implicit solvers and significantly increase computational cost. Therefore, great care must be taken when applying fitting algorithms.

\par \vspace{1em}
\noindent — \textit{Rate Equations} — 
\par
\noindent For transient measurements, the rates of change of the time-dependent carrier populations can be described within the Shockley--Read--Hall (SRH) framework, including an additional radiative recombination term~\cite{shockleyStatisticsRecombinationHoles1952,simmonsNonequilibriumSteadyStateStatistics1971}:
\begin{align}
\frac{\partial n(t)}{\partial t} &= -k_{\mathrm{rad}}\,n(t)\,p(t) - \sum_{i=1}^{n_{\mathrm{traps}}}\left(c_{n,i}(t)-e_{n,i}(t)\right) \label{SI:eq:SRHrateEq1}\\
\frac{\partial p(t)}{\partial t} &= -k_{\mathrm{rad}}\,n(t)\,p(t) - \sum_{i=1}^{n_{\mathrm{traps}}}\left(c_{p,i}(t)-e_{p,i}(t)\right) \label{SI:eq:SRHrateEq2}\\
\frac{\partial n_{\mathrm{tr},i}(t)}{\partial t} &= \left(c_{n,i}(t)-e_{n,i}(t)\right)-\left(c_{p,i}(t)-e_{p,i}(t)\right) \label{SI:eq:SRHrateEq3}
\end{align}

where the capture and emission rates are given by
\begin{equation}
\begin{aligned}
c_{n,i}(t) &= \beta_{n,i}\,n(t)\left(N_{\mathrm{t},i}-n_{\mathrm{tr},i}(t)\right), \qquad e_{n,i}(t) = \beta_{n,i}\,n_{1,i}\,n_{\mathrm{tr},i}(t),\\
c_{p,i}(t) &= \beta_{p,i}\,p(t)\,n_{\mathrm{tr},i}(t), \qquad e_{p,i}(t) = \beta_{p,i}\,p_{1,i}\left(N_{\mathrm{t},i}-n_{\mathrm{tr},i}(t)\right).
\end{aligned}
\label{SI:eq:SRHcaptureEmission}
\end{equation}

Here, \(c_{n,i}(t)\), \(e_{n,i}(t)\), \(c_{p,i}(t)\), and \(e_{p,i}(t)\) denote the electron capture, electron emission, hole capture, and hole emission rates associated with the (i)-th trap species, respectively. The quantities \(\beta_{n,i}\) and \(\beta_{p,i}\) are the corresponding electron and hole capture coefficients, and \(N_{\mathrm{t},i}\) is the density of the (i)-th trap species. The quantities \(n(t), p(t)\), and \(n_{\mathrm{tr},i}(t)\) denote the free-electron density, free-hole density, and density of occupied (i)-th trap states, respectively.

The characteristic carrier concentrations associated with the (i)-th trap energy are
\begin{align}
n_{1,i} &= N_c \exp\left(-\frac{E_c-E_{\mathrm{t},i}}{k_{\mathrm{B}}T}\right), \label{SI:eq:n1}\\
p_{1,i} &= N_v \exp\left(-\frac{E_{\mathrm{t},i}-E_v}{k_{\mathrm{B}}T}\right), \label{SI:eq:p1}
\end{align}
where \(N_c\) and \(N_v\) are the effective densities of states in the conduction and valence bands, respectively. The quantities \(n_{1,i}\) and \(p_{1,i}\) correspond to the equilibrium electron and hole concentrations obtained when the Fermi level coincides with the (i)-th trap energy \(E_{\mathrm{t},i}\).

\par \vspace{1em}
\noindent — \textit{Diffusion term} — 
\par
\noindent As shown in \cref{fig:SI:trPL}, accurate early-time dynamics are of particular importance for distinguishing perovskite thin film samples. In particular, the fluence dependence of these early-time signatures cannot be explained by diffusion-only processes and is instead indicative of fast trapping.

However, carrier diffusion is expected to occur on timescales of \(\sim \SIrange{10}{100}{\nano\second}\)\cite{kober-czernyDeterminingParametersMetalHalide2025,kirchartzPicturingChargeCarrier2022,herzChargeCarrierMobilitiesMetal2017}. Therefore, the inclusion of a diffusion term is necessary to accurately model the system and to disentangle diffusion from other concurrent processes. 

The spatial coordinate $z$ (normal to the film) is therefore introduced and diffusion terms are added to the rate equations
\begin{align}
    \frac{\partial n(t,z)}{\partial t} &= -k_{\mathrm{rad}}\,n(t,z)\,p(t,z) - \sum_{i=0}^{n_{\text{traps}}}\bigl(c_{n,i}(t,z)-e_{n,i}(t,z)\bigr) + D_n \frac{\partial^{2} n(t,z)}{\partial z^{2}} \label{ch5:equation:diffusionelectrons}\\
    \frac{\partial p(t,z)}{\partial t} &= -k_{\mathrm{rad}}\,n(t,z)\,p(t,z) - \sum_{i=0}^{n_{\text{traps}}}\bigl(c_{p,i}(t,z)-e_{p,i}(t,z)\bigr) + D_p \frac{\partial^{2} p(t,z)}{\partial z^{2}} \label{ch5:equation:diffusionholes}\\
    \frac{\partial n_{tr,0}(t,z)}{\partial t} &= \bigl(c_{n,0}(t,z)-e_{n,0}(t,z)\bigr) - \bigl(c_{p,0}(t,z)-e_{p,0}(t,z)\bigr)\\
    &\cdots \notag\\ \label{ch5:equation:diffusiontrapped}
    \frac{\partial n_{tr,i}(t,z)}{\partial t} &= \bigl(c_{n,i}(t,z)-e_{n,i}(t,z)\bigr) - \bigl(c_{p,i}(t,z)-e_{p,i}(t,z)\bigr)
\end{align}

where $D_n$ and $D_p$ [\si{\centi\meter\squared\per\second}] are the electron and hole diffusion constants, respectively.

\par \vspace{1em}
\noindent — \textit{Initial conditions} — 
\par
\noindent 
To calculate the initial conditions for the carrier densities we use the Beer–Lambert generation profile
\begin{align}
    G(z) = \alpha \,N_0\, e^{-\alpha z} &= \alpha\, \frac{F_0}{p_w}\, e^{-\alpha z}\\
    n(t = 0, z) = p(t = 0, z) = \int G(z)\,dt \approx G(z)\,p_w &= \alpha\,F_0\,e^{-\alpha z}\label{ch5:equation:beerlambert:n0}
\end{align}
where $N_0 = \frac{F_0}{p_w}$ [\si{\per\centi\meter\squared\per\second}] is the incident photon flux and $F_0$ [\si{\per\centi\meter\squared}] is the photon fluence, $p_w$ [\si{\second}] is the pulse width and $\alpha$ is the absorption coefficient at the laser excitation wavelength. In \cref{ch5:equation:beerlambert:n0}, we implicitly assume that the background doping ($n_0$, $p_0$) [\si{\per\centi\meter\cubed}] is much lower than the injected carrier density, such that $n(t=0, z) \gg n_0, p_0$.

Finally, the initial conditions for the filling of traps at room temperature are, using the SRH equilibrium conditions \cite{shockleyStatisticsRecombinationHoles1952}
\begin{align} 
    p_{1,i} &= N_v\, e^{\frac{-(E_{\mathrm{t},i}-E_v)}{kT}}\label{ch5:equations:initialconditions1}\\
    n_{1,i} &= N_c\, e^{\frac{-(E_c-E_{\mathrm{t},i})}{kT}}\label{ch5:equations:initialconditions2}\\
    n_i &= \sqrt{N_c N_v}\,e^{\frac{-(E_c-E_v)}{kT}}\label{ch5:equations:initialconditions3}\\
    f_{\mathrm{t},i} &= \frac{\beta_{\mathrm{n},i}\,n_i + \beta_{\mathrm{p},i}\,p_{1,i}}{\beta_{\mathrm{n},i}\,(n_i+n_{1,i}) + \beta_{\mathrm{p},i}\,(n_i+p_{1,i})}\label{ch5:equations:initialconditions4}\\
    n_{\mathrm{t},i} &= f_{\mathrm{t},i}\,N_{\mathrm{t},i} \label{ch5:equations:initialconditions5}
\end{align}

where $E_{\mathrm{t},i}$ [\si{\electronvolt}] is the $i$th trap energy, $E_c$ and $E_v$ [\si{\electronvolt}] are the conduction and valence band energies. $f_{\mathrm{t},i}$ represents the filling probability in the dark at room temperature. $N_{\mathrm{t},i}$ [\si{\per\centi\meter\cubed}] is the $i$th trap density. $n_i$ [\si{\per\centi\meter\cubed}] is the intrinsic carrier density in the bands. $n_{1,i}$ and $p_{1,i}$ [\si{\per\centi\meter\cubed}] are density of trapped electrons or holes in the $i$-th trap, respectively, if the Fermi level were located at the trap energy ($E_f = E_{\mathrm{t},i}$). 

\par \vspace{1em}
\noindent — \textit{Boundary conditions} — 
\par
\noindent We impose zero-flux (Neumann) boundary conditions at both interfaces of the film. The boundary condition is reflected in the calculation of the diffusion term of \cref{ch5:equation:diffusionelectrons} and \cref{ch5:equation:diffusionholes}. We use a central finite-difference scheme in the interior, with boundary expressions consistent with Neumann boundary conditions

\begin{equation}
\frac{\partial^2 n(z,t)}{\partial z^2}
\approx d^2 n_k =
\begin{cases}
\displaystyle \frac{n_{k+1} - 2n_k + n_{k-1}}{\Delta z^2}, & 1 \le k \le N_z-2 \\[10pt]
\displaystyle \frac{2\,(n_1 - n_0)}{\Delta z^2}, & k = 0 \\[10pt]
\displaystyle \frac{2(n_{N_z-2} - n_{N_z-1})}{\Delta z^2}, & k = N_z - 1 \,.
\end{cases}
\end{equation}

\par \vspace{1em}
\noindent — \textit{Jacobian of the system} — 
\par
\noindent 
The stiffness of the coupled recombination--diffusion system requires the use of an implicit integrator (BDF or Radau, see implementations in \cite{virtanenSciPy10Fundamental2020}), which in turn benefits from an analytical Jacobian $J_{ij} = \partial F_i / \partial y_j$ of the right-hand side with respect to the state vector $\mathbf{y}(t,z) = \bigl(n,\ p,\ n_{\mathrm{t},0},\ \ldots,\ n_{\mathrm{t},n_\text{traps}}\bigr)$ at every spatial grid point. Supplying the Jacobian in analytical form, rather than letting the solver estimate it by finite differences, substantially reduces the computational cost of each time step and improves the robustness of the Newton iteration used by the implicit integrator. This efficiency gain is what makes iterative fitting of the model to experimental data feasible within reasonable times.

Denoting the right-hand sides of \cref{ch5:equation:diffusionelectrons,ch5:equation:diffusionholes,ch5:equation:diffusiontrapped} by $F_n$, $F_p$, and $F_{\mathrm{t},i}$ respectively, so that

\begin{equation}
    \frac{\partial n}{\partial t} = F_n,
    \qquad
    \frac{\partial p}{\partial t} = F_p,
    \qquad
    \frac{\partial n_{\mathrm{t},i}}{\partial t} = F_{\mathrm{t},i},
\end{equation}

the non-zero Jacobian blocks of the model are then as follows.

\par \vspace{1em}
\noindent\textit{Free-electron equation.}

\begin{align}
    \frac{\partial F_{n}}{\partial n}
        &= -k_{\mathrm{rad}}\,p
           \;-\; \sum_{i=0}^{n_\text{traps}} \beta_{\mathrm{n},i}\bigl(N_{\mathrm{t},i} - n_{\mathrm{t},i}\bigr)
           \;+\; D_n\,\frac{\partial^{2}}{\partial z^{2}}, \\[4pt]
    \frac{\partial F_{n}}{\partial p}
        &= -k_{\mathrm{rad}}\,n, \\[4pt]
    \frac{\partial F_{n}}{\partial n_{\mathrm{t},i}}
        &= \beta_{\mathrm{n},i}\bigl(n + n_{1,i}\bigr).
\end{align}

\par \vspace{1em}
\noindent\textit{Free-hole equation.}

\begin{align}
    \frac{\partial F_{p}}{\partial p}
        &= -k_{\mathrm{rad}}\,n
           \;-\; \sum_{i=0}^{n_\text{traps}} \beta_{\mathrm{p},i}\,n_{\mathrm{t},i}
           \;+\; D_p\,\frac{\partial^{2}}{\partial z^{2}}, \\[4pt]
    \frac{\partial F_{p}}{\partial n}
        &= -k_{\mathrm{rad}}\,p, \\[4pt]
    \frac{\partial F_{p}}{\partial n_{\mathrm{t},i}}
        &= -\beta_{\mathrm{p},i}\bigl(p + p_{1,i}\bigr).
\end{align}

\par \vspace{1em}
\noindent\textit{Trap equation for species $i$.}

\begin{align}
    \frac{\partial F_{\mathrm{t},i}}{\partial n}
        &= \beta_{\mathrm{n},i}\bigl(N_{\mathrm{t},i} - n_{\mathrm{t},i}\bigr), \\[4pt]
    \frac{\partial F_{\mathrm{t},i}}{\partial p}
        &= -\beta_{\mathrm{p},i}\,n_{\mathrm{t},i}, \\[4pt]
    \frac{\partial F_{\mathrm{t},i}}{\partial n_{\mathrm{t},i}}
        &= -\beta_{\mathrm{n},i}\bigl(n + n_{1,i}\bigr) \;-\; \beta_{\mathrm{p},i}\bigl(p + p_{1,i}\bigr).
\end{align}

\par \vspace{1em}
\noindent — \textit{Spatial discretisation} — 
\par
\noindent 
The diffusion operator $D\,\partial_z^{\,2}$ is discretised on a uniform grid of $N_z$ points with spacing $\Delta z$. Zero-flux (Neumann) boundary conditions at the film interfaces are enforced via a ghost-node construction, yielding

\begin{align}
    \left.D\,\frac{\partial^{2} y}{\partial z^{2}}\right|_k
    &\approx \frac{D}{\Delta z^{2}}\bigl(y_{k-1} - 2 y_k + y_{k+1}\bigr),
    \qquad k = 1, \ldots, N_z - 2 ,\\[4pt]
    \left.D\,\frac{\partial^{2} y}{\partial z^{2}}\right|_0
    &\approx \frac{2 D}{\Delta z^{2}}\bigl(y_1 - y_0\bigr),
    \qquad
    \left.D\,\frac{\partial^{2} y}{\partial z^{2}}\right|_{N_z-1}
    \approx \frac{2 D}{\Delta z^{2}}\bigl(y_{N_z-2} - y_{N_z-1}\bigr).
\end{align}

These contributions populate the tridiagonal sub-blocks of $J$ associated with the $n$ and $p$ rows.

\par \vspace{1em}
\noindent — \textit{Implementation} — 
\par
\noindent The model was initially implemented in a separate codebase, with custom wrappers interfacing with the (now archived) BOAR library~\cite{vincentm.lecorreBOARBayesianOptimization}. It was subsequently integrated into the optimPV library~\cite{vincentm.lecorreOptimPVOptimizationModeling}, which provides a framework for a range of numerical optimisation methods. Notably, the code includes the implementation of an analytical expression for the Jacobian of our system of differential equations.

\section{Comparison with and without Auger terms}\label{SI:Auger}

To assess the possible contribution of direct Auger-Meitner recombination to our decay
dynamics, we modify the continuity equations:

\begin{equation}
\begin{split}
\frac{\partial n(z,t)}{\partial t}
&= -k_{\mathrm{rad}}\, n(z,t)\, p(z,t)
- \sum_{i=0}^{\mathrm{traps}}
  \bigl(c_{e,i}(z,t)-e_{e,i}(z,t)\bigr)
+ D_n \frac{\partial^2 n(z,t)}{\partial z^2} \\
&\quad
- C_{\mathrm{Auger}}
  \bigl(n^2(z,t)\,p(z,t)
  + n(z,t)\,p^2(z,t)\bigr)
\end{split}
\end{equation}

\begin{equation}
\begin{split}
\frac{\partial p(z,t)}{\partial t}
&= -k_{\mathrm{rad}}\, n(z,t)\, p(z,t)
- \sum_{i=0}^{\mathrm{traps}}
  \bigl(c_{h,i}(z,t)-e_{h,i}(z,t)\bigr)
+ D_p \frac{\partial^2 p(z,t)}{\partial z^2} \\
&\quad
- C_{\mathrm{Auger}}
  \bigl(n^2(z,t)\,p(z,t)
  + n(z,t)\,p^2(z,t)\bigr)
\end{split}
\end{equation}

\begin{equation}
\frac{d n_{t,i}(z,t)}{d t} = \bigl(c_{e,i}(z,t) - e_{e,i}(z,t)\bigr)
- \bigl(c_{h,i}(z,t) - e_{h,i}(z,t)\bigr)
\end{equation}

Where $C_{auger}$ is the Auger recombination coefficient. There, we intentionally impose the
electron and hole Auger recombination coefficients to be equal, with an over estimated value of~\cite{staubStatisticsAugerRecombination2018}:

\begin{equation}
C_{auger} = C_n = C_p = \SI{1e-28}{\centi\meter\tothe{6}\per\second}
\end{equation}

With these equations, we then run the simulations with the fitted parameters of the before
irradiation with the CsFAPI sample:

\begin{figure}[h!]
    \centering
    \includegraphics[width=1.0\textwidth]{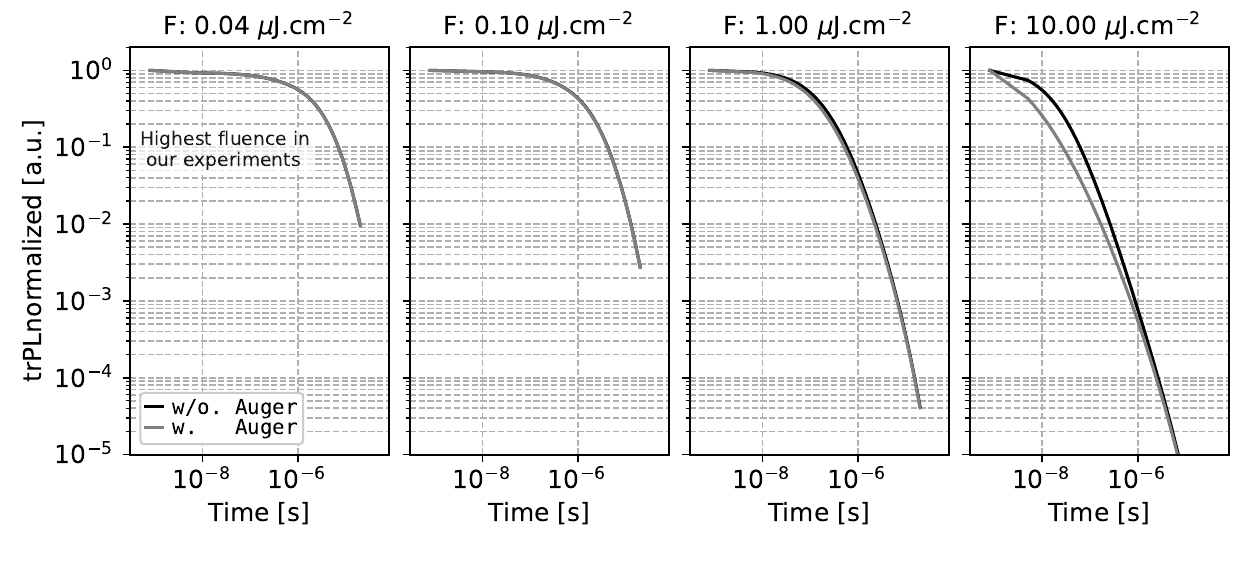}
    \caption{The influence of Auger-Meitner recombination for our trPL decays. Auger-Meitner recombination terms are not included for the black lines. An Auger-Meitner term is included for the grey lines. The parameters correspond to the Cs \SI{15}{\percent} condition. As seen, only difference arise at \(\sim\)100x higher fluences than our maximal experimental value in \cref{fig:SI:trPL}. The Auger coefficient used was \SI{1e-28}{\centi\meter\tothe{6}\per\second}.}
\end{figure}

\newpage
\section{Surface recombination terms}\label{SI:Surfaces}

\begin{figure}[h!]
\centering
\includegraphics[width=1.0\textwidth]{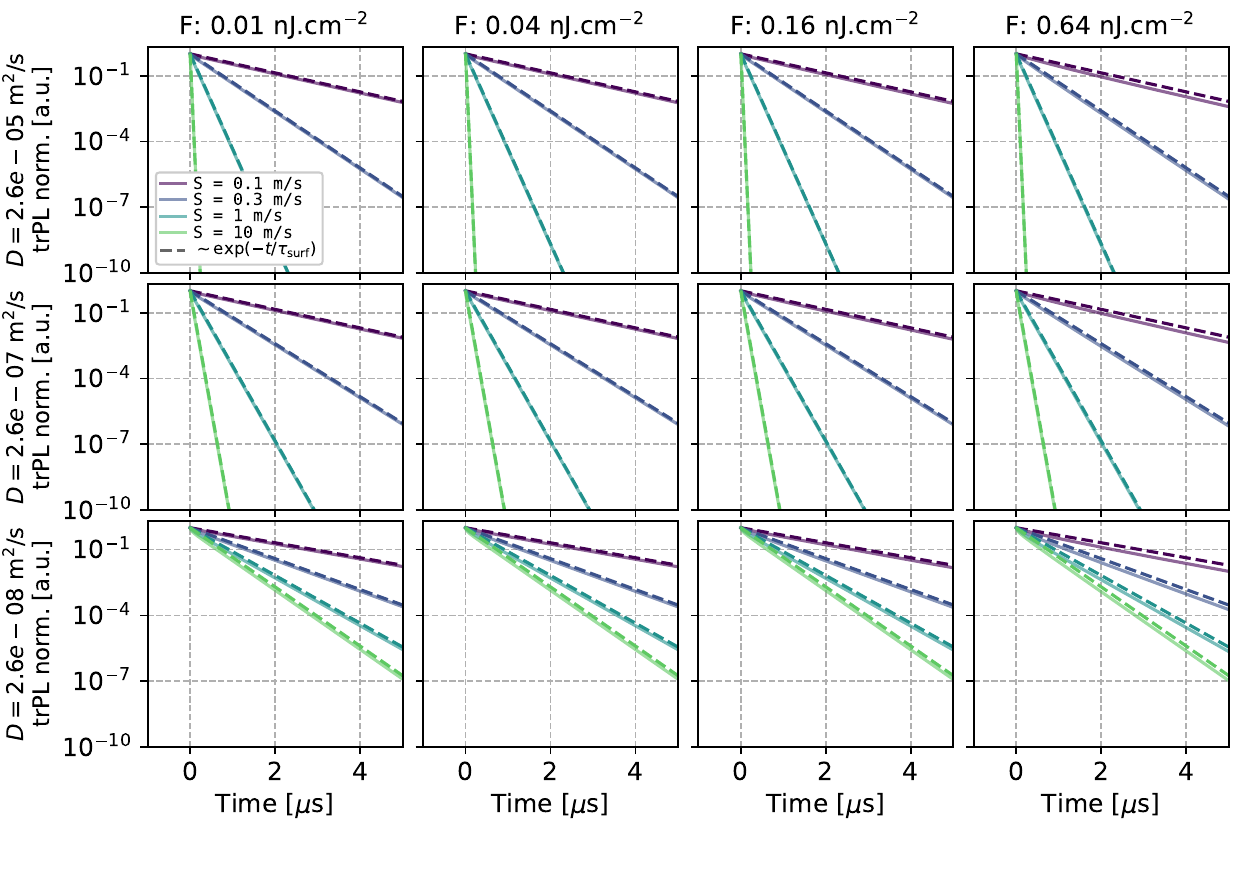}
\caption{\label{SI:fig:Surfaces} Simulated trPL decays with surface recombination as the only carrier-loss pathway. Columns correspond to the four experimental excitation fluences and rows to three carrier diffusion coefficients $D$, spanning the surface-limited to transport-limited regimes. Within each panel, solid curves show the decay for surface recombination velocities $S = 0.1$, $0.3$, $1$, and \SI{10}{\meter\per\second}. Bulk trap-mediated and radiative recombination channels are suppressed. Dashed lines show the corresponding analytical prediction $\mathrm{PL}\propto\exp(-2t/\tau_\mathrm{surf})$, where $\tau_\mathrm{surf}$ is the carrier lifetime obtained from the eigenvalue condition $\tan(\beta L/2) = S/(D\beta)$ with $\tau_\mathrm{surf} = 1/(D\beta^2)$ found in \textcite{lukeAnalysisInteractionLaser1987}; the factor of two arises because the PL signal scales as $n\,p$ and therefore decays at twice the single-carrier rate.}
\end{figure}

For all conditions investigated here, the simulated transients are effectively single-exponential. More generally, a constant surface recombination velocity coupled to linear diffusion produces a superposition of exponential diffusion eigenmodes, with the long-time dynamics dominated by the lowest-order mode\cite{lukeAnalysisInteractionLaser1987}. It therefore does not provide a mechanism for the extended power-law dynamics observed experimentally. In the large-$D$ row the decays are effectively single-exponential and fluence-independent, and match the surface-limited limit $\tau_\mathrm{surf} \to L/(2S)$. As $D$ decreases (lower rows), transport of carriers to the surfaces becomes rate-limiting: the decay slows, loses its sensitivity to $S$, and is set instead by the diffusion time across the film. In all cases the decay remains a single exponential and does not reproduce the power-law dynamics observed experimentally. This justifies the removal of a surface recombination term in order to allow for power-law dynamics to emerge from other terms such as bulk recombination.

\section{Equilibration versus no equilibration}\label{SI:Equilibration}

\begin{figure}[h!]
    \centering
    \includegraphics[width=1.0\textwidth]{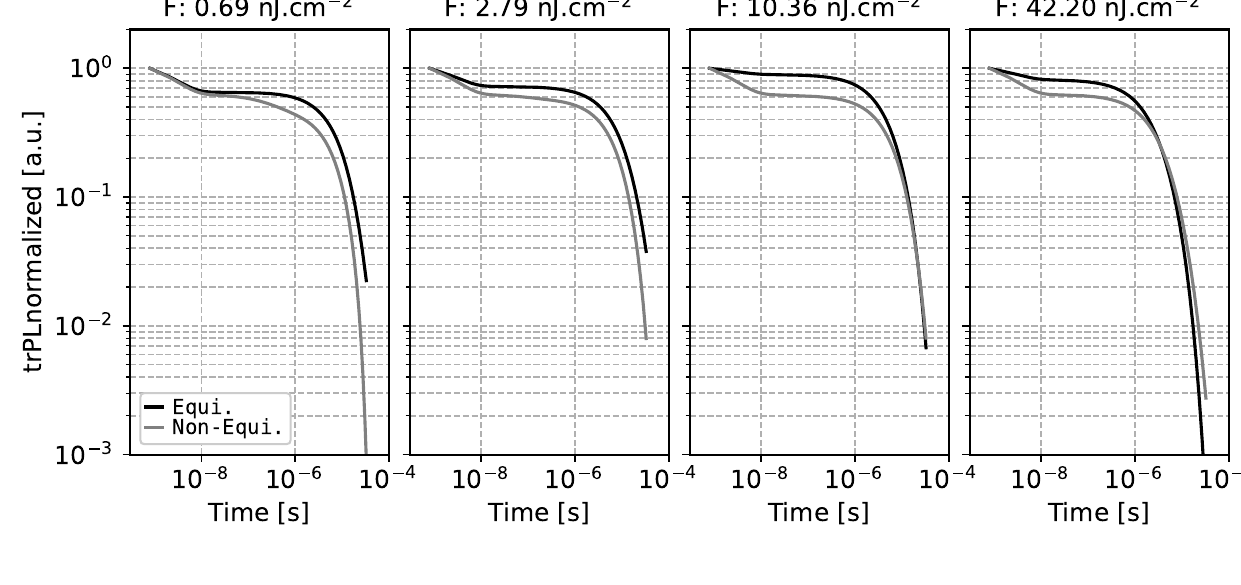}
    \caption{The effect of neglecting trapped-charge equilibration on the transient decay shape (in grey), compared to an equilibrated decay (dark). We note that these differences are function of the simulated parameter set, this example demonstrates that trapped-charge equilibration must be accounted for when fitting over wide parameter ranges.}
\end{figure}

\section{Model Complexity: Selection of number of trapping signatures}\label{SI:ModelComplexity}

\begin{figure}[h!]
  \centering
  \includegraphics[width=1\columnwidth]{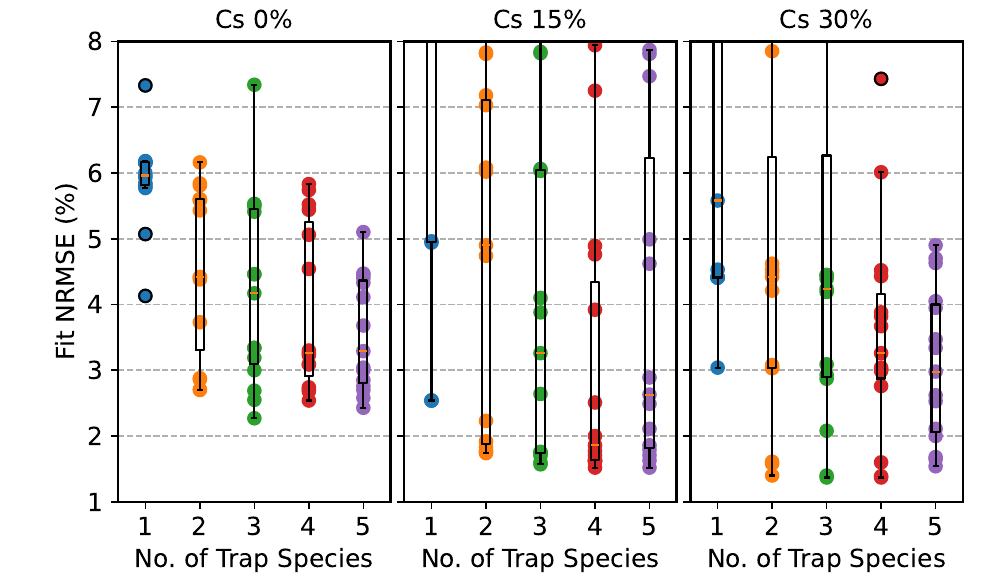}
  \caption{\label{fig:SI:fit_statistics}Dependence of the global fit quality on the number of trapping species for the Cs0, Cs15, and Cs30 films. Bayesian optimisation was repeated fifteen times for each model complexity. The selected models contain $i=3$ species for Cs0, Cs15, and Cs30.}
\end{figure}

The number of trapping species, $i$, required to describe the fluence-dependent trPL measurements was first assessed independently for each composition by systematically increasing the model complexity. For each value of $i$, Bayesian optimisation was repeated fifteen times from independent initial conditions to reduce sensitivity to local optima. We selected the smallest model beyond which adding an additional trapping species did not yield a consistent reduction in the best-fit NRMSE. Across the compositional series, three trapping species were sufficient to describe the measured dynamics. We therefore use $i=3$ for all compositions to maintain a consistent model dimensionality and facilitate direct comparison of the inferred trapping signatures. As shown in \cref{fig:SI:fit_statistics}, increasing the Cs30 model beyond two trapping species does not yield significant improvement in NRMSE. For Cs15 and Cs00, this happens at 3 defect species.

\begin{figure}[h!]
  \centering
  \includegraphics[width=1\columnwidth]{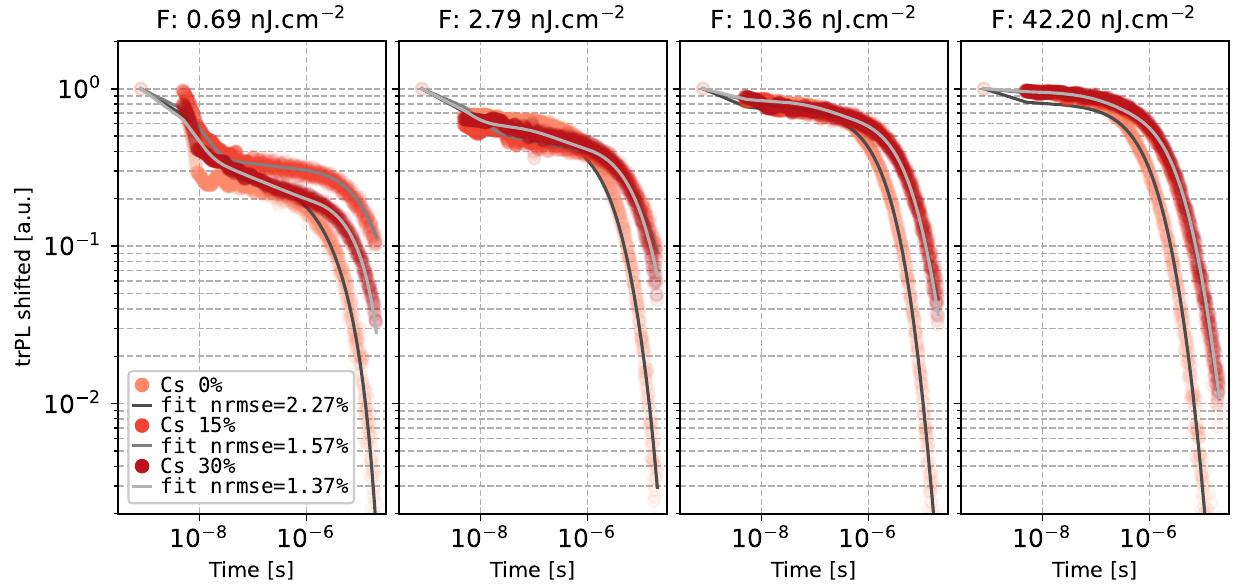}
  \caption{\label{fig:SI:TuRBO_fits}Global fits to the fluence-dependent trPL measurements using the selected models, with $n_t=3$ for all the Cs0, Cs15, and Cs30 films, respectively.}
\end{figure}

The corresponding fits for the best fit values are shown in \cref{fig:SI:TuRBO_fits}, for $i=3$.

\section{Full Corner Plots}\label{SI:FullCornerPlots}

\begin{figure}[H]
  \centering
  \includegraphics[width=1.05\columnwidth]{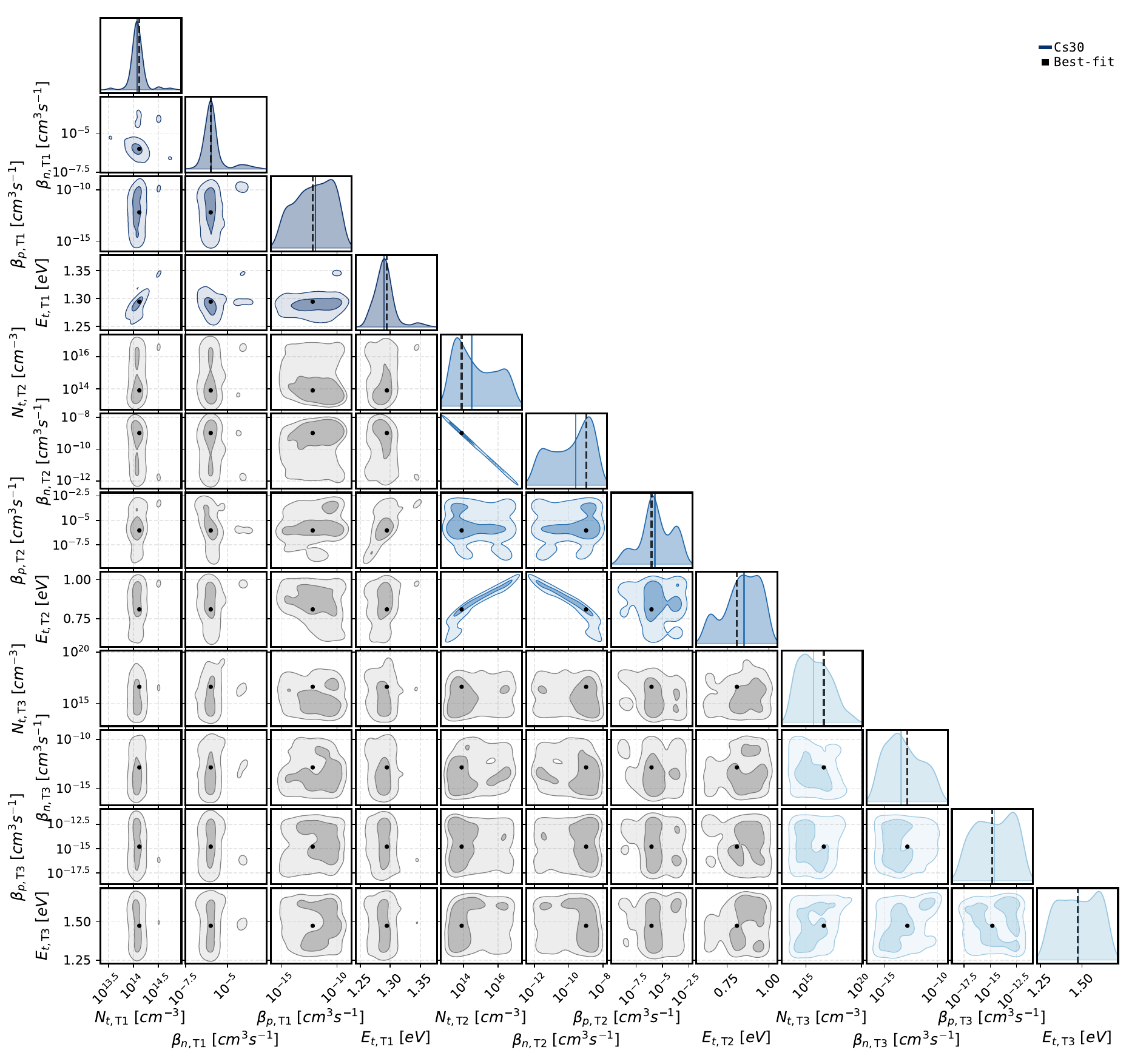}
  \caption{Posterior distributions and pairwise parameter correlations for all trapping parameters of the Cs30 film. We do not observe pronounced inter-signature correlations, for example between parameters associated with T1 and T2. The coloured pairwise distributions correspond to the intra-signature correlations shown in the main text.}
\end{figure}

\begin{figure}[H]
  \centering
  \includegraphics[width=1.05\columnwidth]{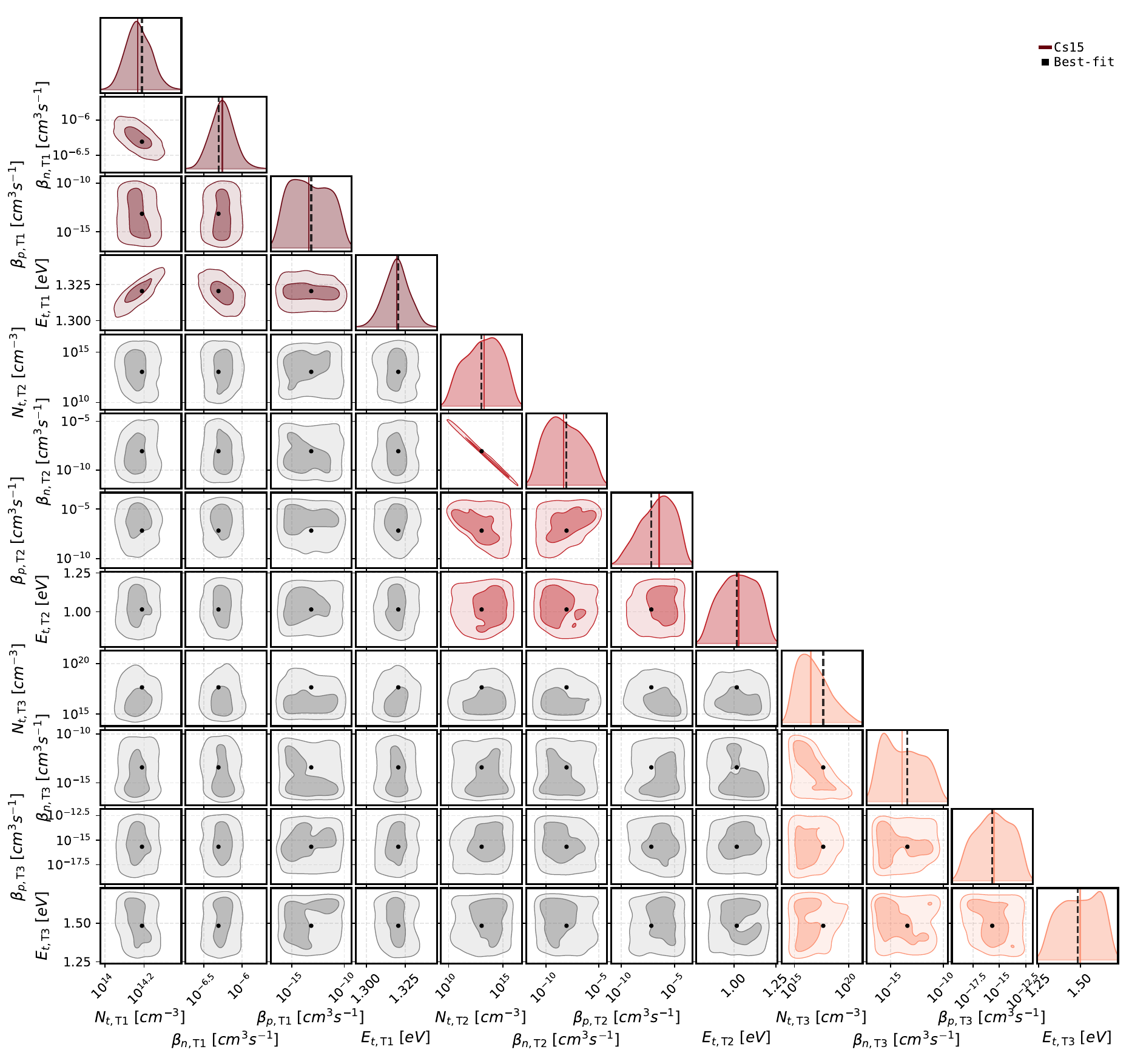}
  \caption{Posterior distributions and pairwise parameter correlations for all trapping parameters of the Cs15 film. We do not observe pronounced inter-signature correlations, for example between parameters associated with T1 and T2. The coloured pairwise distributions correspond to the intra-signature correlations shown in the main text.}
\end{figure}

\begin{figure}[H]
  \centering
  \includegraphics[width=1.05\columnwidth]{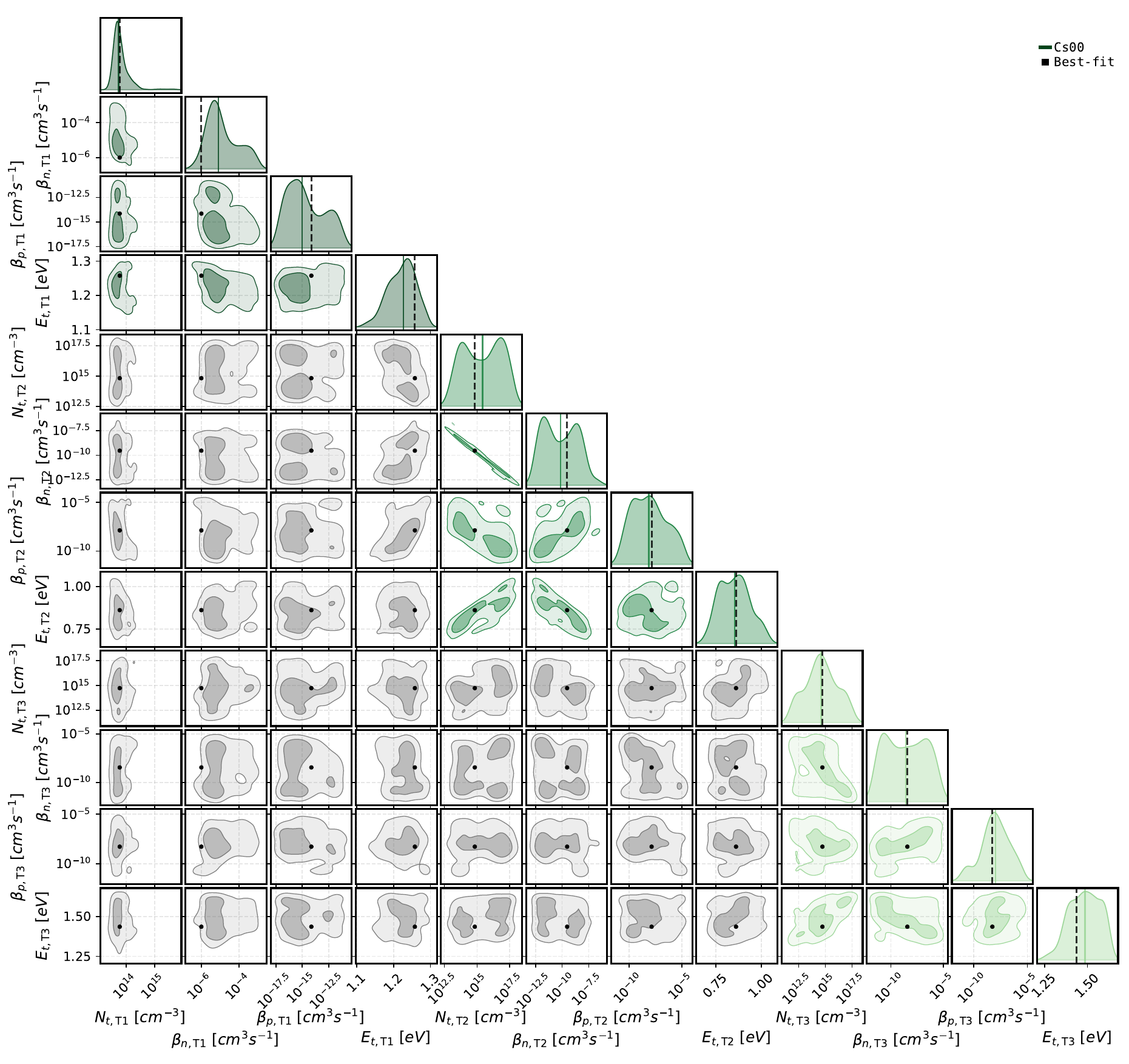}
  \caption{Posterior distributions and pairwise parameter correlations for all trapping parameters of the Cs00 film. We do not observe pronounced inter-signature correlations, for example between parameters associated with T1 and T2. The coloured pairwise distributions correspond to the intra-signature correlations shown in the main text.}
\end{figure}

\section{\texorpdfstring{\(T\beta\) displaying \(\sqrt{np} \propto n^{-1}\)}{T beta displaying sqrt(np) proportional to n^-1}}
\label{SI:1overN}

\begin{figure}[H]
\centering
\includegraphics[width=0.6\textwidth]{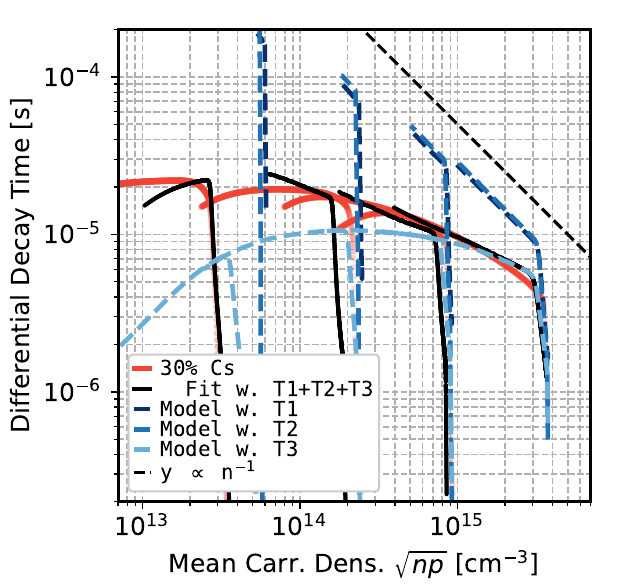}
\caption{\label{SI:fig:1overN}Differential decay time in function of average carrier density for the Cs30 sample. The original data is plotted in red, the fit containing both trapping species is in black. The dashed dark blue curve is the resulting modeled curve only keeping the first trapping species (T2). T2 shows pronounced $\beta_n$-$N_t$ degeneracy in \cref{fig:fig3:MCMC-Cs30} which results in a flat differential decay time curve, consistent with more traditional trapping signatures of inorganic semiconductors. On the other hand, T1 displays no trap degeneracy, which we show is consistent with differential decay time slope $y \propto n^{-1}$. Such $\propto n^{-1}$ regime has previously been predicted when trap-mediated electron capture and emission are balanced~\cite{yuanUnderstandingPowerLawPhotoluminescence2024}.}
\end{figure}

As shown in \cref{SI:fig:1overN}, T1 associated with this $1/n$ dependence. In the main text, it exhibits substantial trap occupation and a lifted $\beta_n\,N_t$ degeneracy. These observations connect the experimentally observed $1/n$ regime with significant trap filling, which is compatible with balanced electron capture and emission rates,
\begin{equation}
\begin{aligned}
    c_{n}(t)
    &= \beta_{n}\,\bigl(N_{t}-n_{tr}(t)\bigr)\,n(t), \\
    e_{n}(t)
    &= \beta_{n}\,N_c\,\exp\!\left(-\frac{E_c-E_{t}}{kT}\right)\,n_{tr}(t).
\end{aligned}
\label{eq:cn_en}
\end{equation}
Defining
\begin{equation}
    n_{1}
    = N_c\,\exp\!\left(-\frac{E_c-E_{t}}{kT}\right),
\end{equation}
the condition $c_{n}(t)\simeq e_{n}(t)$ yields
\begin{align}
    \bigl(N_{t}-n_{tr}(t)\bigr)\,n(t)
    \simeq n_{1}\,n_{tr}(t),
    \label{eq:capture_emission_balance}
\end{align}
as described in \textcite{yuanUnderstandingPowerLawPhotoluminescence2024}. The occurrence of this regime for T2 is therefore consistent with significant trap occupation, as suggested by its lifted $\beta_n\,N_t$ degeneracy.

\subsection{Steady state recombination}

\begin{figure}[H]
\centering
\includegraphics[width=0.98\textwidth]{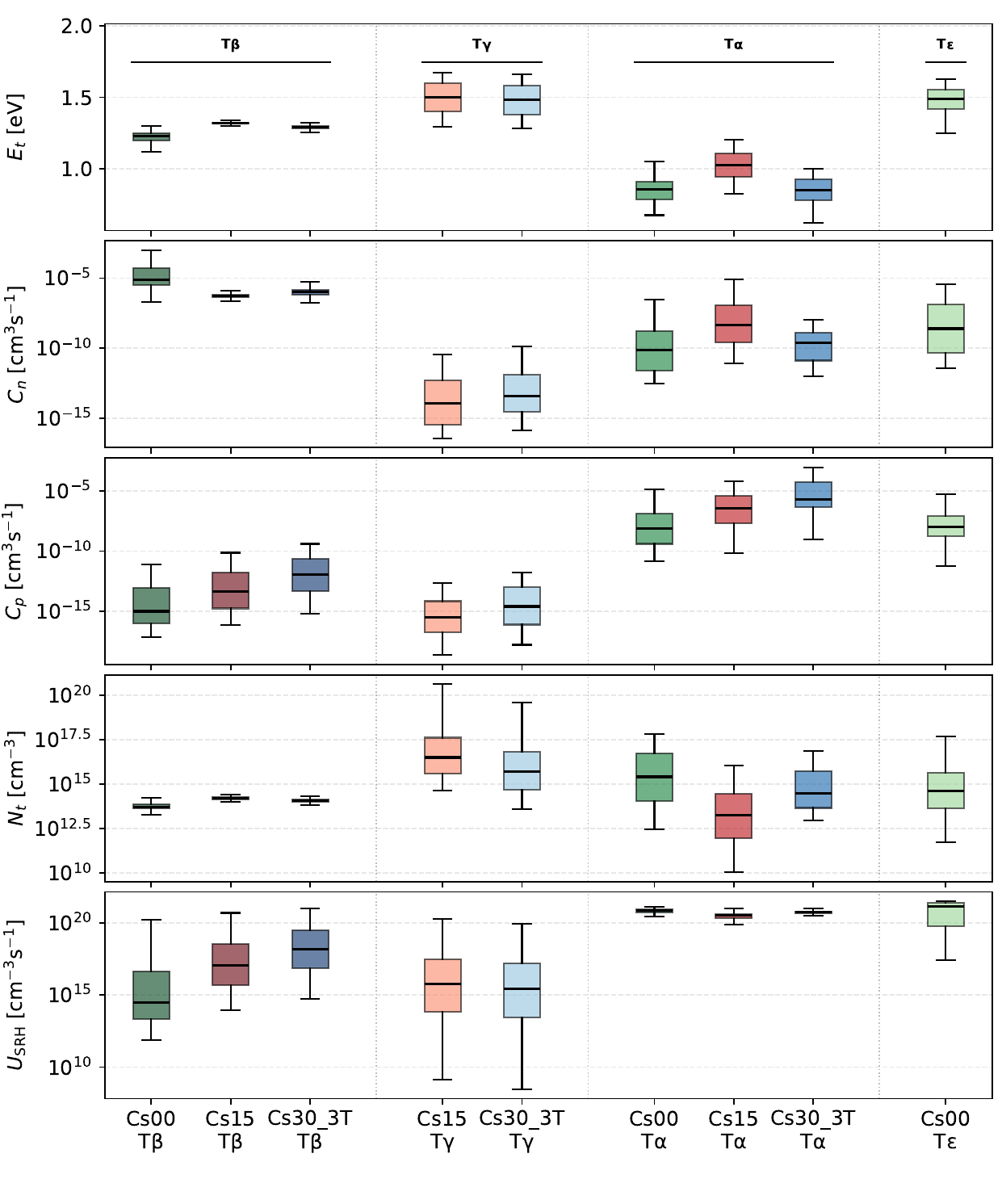}
\caption{\label{fig:SI:MCMC-Boxplot}Summary of the inferred parameter distributions for the four effective trapping signatures identified across the FA$_{1-x}$Cs$_x$PbI$_3$ compositional series, together with their corresponding steady-state SRH recombination rates.}
\end{figure}

For estimation of the recombination properties associated with the inferred trapping parameters, we assume a constant photon flux of \(N_0 = \SI{1.5e17}{\per\centi\meter\squared\per\second}\), corresponding approximately to a 1-Sun photon flux for a bandgap of \SI{1.55}{\electronvolt}. For each posterior parameter set, the steady-state carrier densities are determined and the corresponding defect-specific Shockley--Read--Hall recombination rate \(U^{\mathrm{SS}}_{\mathrm{SRH},i}\) is evaluated for each trapping signature. This allows us to assess which signature contribute most strongly to non-radiative recombination under solar-cell operating conditions.

For trapping signature \(i\), the SRH recombination rate is calculated according to~\cite{shockleyStatisticsRecombinationHoles1952}
\begin{align}
U_{\mathrm{SRH},i} =
N_{t,i}\,
\frac{\beta_{n,i}\beta_{p,i}\left(np-n_i^2\right)}
{\beta_{n,i}\left(n+n_{1,i}\right)+
\beta_{p,i}\left(p+p_{1,i}\right)}.
\label{eq:SS-SRH}
\end{align}

The steady-state carrier densities \(n\) and \(p\) are obtained numerically by simultaneously solving the charge-neutrality condition for the occupied electron traps,
\begin{equation}
p-n-\sum_{i=1}^{n_{\mathrm{traps}}} n_{\mathrm{tr},i}=0,
\label{eq:eqUSRH_chargeNeutr}
\end{equation}
and the total generation--recombination balance,
\begin{equation}
G-R_{\mathrm{rad}}-\sum_{i=1}^{n_{\mathrm{traps}}}U_{\mathrm{SRH},i}=0,
\label{eq:eqUSRH_Gen}
\end{equation}
where the radiative recombination rate is
\begin{equation}
R_{\mathrm{rad}} = k_{\mathrm{rad}}np.
\end{equation}

Thus, although \(U^{\mathrm{SS}}_{\mathrm{SRH},i}\) denotes only the non-radiative SRH contribution of an individual signature, its steady-state value is evaluated using carrier densities that result from competition between radiative recombination and the SRH recombination of all signatures.

\section{Parameter Inference}\label{SI:Usrh}
Here, we use a global fitting approach that aims to infer a single set of material parameters $\mathcal{P}$ from a series of 4 trPL decay curves measured on the same sample:

\begin{equation}
\mathcal{P} =
\left\{
E_g,\; L,\; \alpha,\; N_{cv},\;
k_{\mathrm{direct}},\;
\mu_n,\; \mu_p,\;
\left( N_{t,i},\, \beta_{n,i},\, \beta_{p,i},\, E_{t,i} \right)_{i=1}^{n_t}
\right\}
\end{equation}

where $n_t$ is the number of trapping species considered in the model. Within $\mathcal{P}$, a subset of parameters is treated as fixed, either determined independently ($E_g,\; L,\; \alpha$) or taken from the literature ($N_{cv}$):

\begin{equation}
\mathcal{P}_{\mathrm{fixed}} =
\left\{
E_g,\; L,\; \alpha,\; N_{cv}
\right\}.
\end{equation}

The remaining parameters are treated as unknown and are inferred through fitting

\begin{equation}
\mathcal{P}_{\mathrm{unknown}} =
\left\{
k_{\mathrm{direct}},\;
\mu_n,\; \mu_p,\;
\left( N_{t,i},\, \beta_{n,i},\, \beta_{p,i},\, E_{t,i} \right)_{i=1}^{n_t}
\right\}.
\end{equation}

The total number of unknown parameters, $n_{p,\mathrm{unknown}} = 3 + 4\,n_t$, therefore depends strongly on the assumed number of trapping species.

\subsection{Analytical expression for the dependence of trap filling and beta ratio}

\bibliographystyle{apsrev4-2}
\bibliography{test}

\end{document}